\documentclass[pdflatex,sn-mathphys-ay]{sn-jnl}

\usepackage{graphicx}%
\usepackage{multirow}%
\usepackage{amsmath,amssymb,amsfonts}%
\usepackage{amsthm}%
\usepackage{mathrsfs}%
\usepackage[title]{appendix}%
\usepackage{xcolor}%
\usepackage{textcomp}%
\usepackage{manyfoot}%
\usepackage{booktabs}%
\usepackage{algorithm}%
\usepackage{algorithmicx}%
\usepackage{algpseudocode}%
\usepackage{listings}%
\usepackage{adjustbox}
\usepackage[T5]{fontenc}
\usepackage{caption}
\usepackage{multicol, multirow}
\usepackage{float}  
\usepackage{subcaption}
\usepackage{adjustbox}
\usepackage{longtable}
\theoremstyle{thmstyleone}%
\theoremstyle{thmstyletwo}%

\theoremstyle{thmstylethree}%

\begin{document}

\title[Article Title]{Large Language Models in Teaching and Learning: Reflections on Implementing an AI Chatbot in Higher Education}

\author*[1]{\fnm{Fiammetta} \sur{Caccavale}}\email{fiacac@dtu.dk}

\author[1]{\fnm{Carina L.} \sur{Gargalo}}

\author[1]{\fnm{Julian} \sur{Kager}}

\author[1]{\fnm{Magdalena} \sur{Skowyra}}

\author[1]{\fnm{Steen} \sur{Larsen}}

\author[1]{\fnm{Krist V.} \sur{Gernaey}}

\author*[1]{\fnm{Ulrich} \sur{Kr{\"u}hne}}\email{ulkr@kt.dtu.dk}

\affil[1]{\orgdiv{Department of Chemical and Biochemical Engineering}, \orgname{Technical University of Denmark}, \orgaddress{\street{Søltofts Plads 228A}, \city{Kongens Lyngby}, \postcode{2800}, \country{Denmark}}}

\abstract{
The landscape of education is changing rapidly, shaped by emerging pedagogical approaches, technological innovations such as artificial intelligence (AI), and evolving societal expectations, all of which demand thorough evaluation of new educational tools. Although large language models (LLMs) present substantial opportunities especially in Higher Education, their propensity to generate hallucinations and their limited specialized knowledge may introduce significant risks. This study aims to address these risks by examining the practical implementation of an LLM-enhanced assistant in a university level course.

We implemented a generative AI assistant grounded in a retrieval-augmented generation (RAG) model to replicate a previously teacher-led, time-intensive exercise. To assess the effectiveness of the LLM, we conducted three separate experiments through iterative mixed-methods approaches, including a crossover design. The resulting data address central research questions related to student motivation, perceived differences between engaging with the LLM versus a human teacher, the quality of AI-generated responses, and the impact of the LLM on students’ academic performance. The results offer direct insights into students’ views and the pedagogical feasibility of embedding LLMs into specialized courses. Finally, we discuss the main challenges, opportunities and future directions of LLMs in teaching and learning in Higher Education.
}


\maketitle

\section{Introduction}
\label{introduction}

Education is constantly evolving with new pedagogies (e.g., blended learning, competency-based education), technologies (e.g., artificial intelligence, mixed and virtual reality), and societal demands (e.g., 21st-century skills, sustainability education). 
Among these technologies, artificial intelligence (AI) is starting to be consistently integrated in Higher Education. In recent years, researchers have been particularly intrigued large language models (LLMs) and the impact that this technology could have on education. Possible benefits that LLMs could provide include personalized learning, increased teaching efficiency, tailored student support, which could lead to enhanced learning experience and overall higher
engagement of students \citep{kasneci2023chatgpt}. 

However, there are also arguably as many challenges introduced by this technology, such as over-reliance on LLMs \citep{abd2023large}, which could hinder critical thinking and reasoning skills \citep{xu2024opportunities}. These challenges also include the tendency of LLMs to hallucinate, leading to students receiving incorrect information, and to be too generalized to provide domain-specific knowledge that the Higher Education curriculum requires. Other issues include possible ethical concerns, ensuring transparency, privacy and accountability \citep{yan2024practical}.
Therefore, there is a need for educators to scrupulously consider applications and alignmentof LLMs with learning objectives, as well as thoroughly validate these models before they are used in education. If no precautions are taken, these issues could cause the introduction of LLMs in education to inhibit rather than facilitate learning.

The purpose of this study is to contribute to the ongoing research and discussion on the introduction of LLMs in Higher Education. We introduce a chatbot previously developed by the authors \citep{caccavale2025chatgmp}, which is based on a pre-trained LLM using retrieval-augmented generations (RAG), to a Master's Degree course. 
The chatbot is created to simulate an interview-like exercise between the students (asking the questions) and a fictional company (originally represented by the teachers of the course)
An iterative mixed-methods approach is used to validate the developed chatbot. We provide extensive first-hand opinions from students who tested this technology in three different experiments, including a crossover study. 
With the collected data and analysis in this contribution we show: (i) reasons and motivations of students to interact with a chatbot instead of a teacher, (ii) the main differences between interacting with a teacher compared to the chatbot, (iii) the effect on the grading, and (iv) the perceptions of adding speech-to-text and text-to-speech functions on the user experience.

The learning from this experiments 
Finally, we summarize the opportunities and challenges that LLMs can introduce to education, and in Section \ref{discussion}, we highlight the pedagogical challenges associated with AI and present our considerations on how this technology can be leveraged in collaborative learning.

\section{Background}
\label{background}
This section introduces the current background for the introduction of AI models in Higher Education. First, the opportunities that LLMs could bring to this field will be discussed, followed by an introduction to their benefits to collaborative learning. Lastly, current challenges and possible drawbacks and limitations will be presented.

\subsection{Opportunities introduced by LLMs in education}
Current advancements in AI, specifically with regards to large language models (LLMs) offer substantial benefits in Higher Education, including personalized learning \citep{Gan2023Large}, tailored student support \citep{rodrigues2024assessing}, teaching efficiency \citep{Xu2024Large}, greater student engagement \citep{Pelaez-Sanchez2024The}, support for research, and increased accessibility and inclusivity \citep{Gan2023Large, Xu2024Large}.

These benefits can lead to an enhanced learning experience and overall higher engagement of students \citep{kasneci2023chatgpt}, as well as improvements in student comprehension and academic outcomes \citep{rodrigues2024assessing}. 
Research indicates that AI can greatly improve learning efficiency by automatically regulating cognitive load, offering personalized teaching, and dynamically adjusting learning pathways \citep{gkintoni2025challenging}. This research highlights that AI-based strategies are particularly advantageous in high cognitive load domains such as STEM, where they can substantially enhance understanding, memory retention, and problem-solving abilities \citep{gkintoni2025challenging}. This evidence on the efficacy of LLMs as learning aids suggests that the future of engineering education will include LLMs \citep{kamalov2023new}. 

\subsection{LLM support in collaborative learning}
Research has observed that LLMs can support collaborative learning by acting as intelligent conversational agents, providing real-time answers or feedback, and helping students by generating discussions tailored to group activities \citep{naik2024generating, naik2025providing}. For example, LLMs can support group reflection and offer alternative solutions, helping students engage more deeply, both with content and with each other. They can also enable inclusive collaboration by detecting and modeling group dynamics and offer feedback to improve teamwork and communication skills \citep{cai2024advancing, lewis2022multimodal}.

\subsection{Pedagogical challenges introduced by AI} 
Educational research often struggles to keep up with the pace of the changes introduced by new pedagogies, technologies and societal demands, making it difficult to conduct timely and relevant studies and capture the nuances of new learning environments \citep{lavicza2022developing}. Moreover, traditional educational research may not be fully equipped to study AI in education \citep{chiu2024future}. Despite the hype surrounding AI, there is a shortage of evidence for its successful incorporation and validation in real-world educational environments \citep{yan2024practical}.
Studies indicate that most LLM advancements lack a focus on human-centered design and the involvement of educational stakeholders. Consequently, these tools might impede learning rather than improve it \citep{yan2024practical}. 
The integration of innovative technology into education presents a complex challenge \citep{obidovna2024pedagogical}, as learning objectives and teaching activities need meticulous adjustment \citep{aggarwal2023integration}.
Additionally, the lack of transparency in data collection could result in the misuse of student information and an erosion of trust \citep{nguyen2023ethical}. Academic integrity may likewise suffer, as students could employ AI to circumvent learning objectives, thereby diminishing the significance of assessments and complicating the evaluation of genuine learning, creating an unfair environment \citep{kelly2023chatgpt}. 
AI's integration into education also profoundly impacts the role of teachers. Research suggests that educators require comprehensive training to learn the capabilities and limitations of AI, to effectively integrate it into pedagogy, develop innovative assessment strategies, and promote the responsible use of LLMs \citep{nguyen2025use}. Otherwise, AI tools might be misused or not perceived as a valuable asset \citep{ng2023teachers}. 
Effective integration of AI into education requires addressing the current challenges and aligning with educational objectives, careful management, teacher training, ethical considerations and data privacy safeguards \citep{Zhang2024Reflections, Idris2024Revolutionizing, caccavale2024llm}.
Therefore, although the benefits that AI could introduce in education are substantial, it is necessary to review the opportunities, challenges, and future directions that LLMs can generate in teaching and learning.


\section{Methods}
\label{methods}
In this section, we introduce the architecture of the AI assistant presented in this study, as well as the mixed-methods iterative approach used to develop and validate it, including a crossover study to further gather insights into students' interaction and perception of the chatbot.

\subsection{Aim of the exercise}
In this study, we investigate the introduction of a custom LLM in the "Good Manufacturing Practice (GMP) and quality in pharmaceutical, biotech and food industry - Theoretical version" (28855) Master's Degree course. We use a chatbot previously developed by the authors \cite{caccavale2025chatgmp}, which is based on a pre-trained LLM using RAG. 
The exercise is created to simulate an audit in a pharmaceutical company. The purpose of the exercise is for the students to evaluate the company's compliance to standard procedures; practically, they prepare and perform a mock audit to gather evidence, by asking questions and documents. This information is required to determine whether the company under review could be a suitable business partner or whether the identified non-conformities are too serious. Teachers are expected to either provide precise answers when the necessary documents are available or to respond more vaguely, encouraging students to think critically about the company’s conduct. The deliverable for this exercise is an audit report. This is not the final exam, but a group assignment that must be passed to gain access to the exam. This report is assessed to evaluate the groups’ skills in identifying minor and major non-conformities and in linking them to the relevant legislation. In a concluding section, the students must state whether the company is compliant and, if it is not, propose an appropriate action plan.
In previous years, the fictional company was represented by teachers; however, due to the high number of students enrolled in the course (maximum uptake of 120 students; however, 25\% more students apply to enroll) and the few teachers available (3 teachers per semester), this task was considered quite repetitive and time-consuming. Nevertheless, teachers did not want to eliminate it from the course, since it leads to significant learning outcomes. 
Thus, automating this process would benefit both students, increasing the yearly uptake and thus giving the possibility to more students to enroll, and teachers, automating a time-consuming and repetitive task. The course is generally discursive, and the regulations and documentation are straightforward to discuss, so it poses a suitable use case for the application of AI. 
Therefore, the solution was to gradually introduce and validate a chatbot that could substitute the teacher in this exercise.

It should be emphasized that this initiative is not intended to create a course without a teacher. Instead, its purpose is to allow teachers to concentrate on delivering lectures and preparing course materials, rather than spending substantial time on exercises that are highly time-consuming and repetitive. It is also essential to recognize that individual students may have different preferences regarding whether they feel comfortable interacting with a chatbot or prefer engaging with a teacher in person. Therefore, participation in this experiment was entirely voluntary, so that groups can choose whether to carry out the audit with a teacher or with the LLM.

\subsection{LLM-enhanced learning system}

The full model pipeline, including data collection, pre-processing, modelling and deployment, is explained in details in \cite{caccavale2025chatgmp}. 
The virtual assistant implemented and used in this study is a sequence-to-sequence question-answering (QA) model. Students interact with the system through a graphical user interface (GUI). The LLM component is based on a pre-trained, open-source model, which can be downloaded from the Hugging Face library \citep{huggingface}, FLAN-T5 base \citep{chung2024scaling}. A smaller model, compare to the more complex and powerful LLMs currently available, was chosen in order to limit the risk of hallucinations and ensure the correctness of the answers, since the model does not majorly deviates from the provided context.

We use retrieval-augmented generation (RAG), a technique that enables LLMs to retrieve and incorporate new information in the responses \citep{lewis2020retrieval}. Therefore, the prompt is enriched with further context, which is retrieved through semantic search by calculating the cosine similarity between the input question (asked by the student) and all questions contained in the database. In this case, the context is the historical answer provided by a teacher. The LLM then generates an answer to the question asked based on the context provided.

The backend of the model was developed in Python, using the Flask package for deployment. The front-end is in HTML. The model was deployed locally using a single NVIDIA GeForce RTX 2060 with Max-Q Design GPU. During the exercise, students interacted with the virtual assistant through a laptop connected to a monitor. If a specific document was requested by the students, the model was triggered to show the document through a pop-up window. The students then had the opportunity to skim the document before proceeding to the next question. All the documents shown to the students during the audit were later sent to them, together with the transcript of their interaction with the model.

\subsection{Mixed-methods iterative approaches}

This work adopts mixed-methods iterative approaches that combine quantitative data (e.g., performance, metrics from AI systems) with qualitative insights (e.g., student and teacher experiences). To develop the model, various LLMs were benchmarked in previous studies; we report the performance of a selection of the models used for the comparison in the Appendix, in Table \ref{tab:results_appendix}.

Regarding qualitative data, we collected students' perspectives before and after the experiment. 
To evaluate the chatbot, we define four principal research questions (RQ), which are presented below.
\begin{itemize}
    \item RQ1: "How happy are you with your experience?"
    \item RQ2: "After having performed the audit, what do you think of the quality of the answers?"
    \item RQ3: "Would you recommend doing the audit with your auditee (teacher or chatbot) to other students?"
    \item RQ4: "Do you think this is the future of the audit exercise in this course?"
\end{itemize}
RQ1, RQ2 and RQ3 are scored according to a Likert scale in range 1-5, from lowest to highest. RQ4 is a categorical question with three possible given answers: "Yes", "I am not sure" and "No".
The model has been used by students on three occasions: (a) in the spring semester of 2024, with 3/21 groups of students (N=18, total=119) i.e., the chatbot was used with 3 groups of students (6 students each) out of the 21 total groups; (b) in the spring semester of 2025, with 9/23 groups of students (N=43, total=113); and, (c) in a crossover study carried out in the fall semester of 2025 with 2/2 groups (N=6, total=6). Students could voluntarily decide whether to fill in the survey; therefore, we collected responses from: (i) 14 students performing the audit with a teacher and 13 with the AI assistant in 2024; (ii) 13 students performing the audit with a teacher and 18 with AI assistant in 2025; and, (iii) 6 students in the crossover study performing the audit with both the teacher and AI assistant.

For each cohort, to assess whether the difference between the two groups is significant, two statistical tests were performed. 
We conducted a Mann-Whitney U test \citep{mann1947test}, which is a non-parametric test that compares the distributions of two independent groups for the first three research questions (RQ1, RQ2 and RQ3). 
For the last research question (RQ4), we performed a Chi-square test \citep{mchugh2013chi}, which assesses whether there is a significant association between two categorical variables.

\subsection{Crossover study}

To further test the model and gain new insights, we employ the model again in the fall semester of 2025. This time, the model was tested with a different cohort of students, and hence we cannot assume that the results will be comparable, due to the differences in experience and background of this group of students. 

There were six students participating in the crossover study, divided into two groups of three students each. To collect their perspectives on the different interactions and what the strengths and weaknesses are, we asked each group to perform the audit exercise with both the teacher and the chatbot. The experiment lasted a total of 1.5 hours, time allocated equally between the chatbot and the teacher. 
The students were made aware in advance of the use of AI in this study, and informed consent was collected before starting the experiment. As in the previous experiment, the interaction was not evaluated; only the report of the findings is graded.

\section{Results}
\label{results}

In this Section, the results of the three experiments, together with the aggregated grades of the students participating in this initiative, will be presented. Firstly, we will compare how students' perspectives have shifted between 2024, when the chatbot was originally introduced, and 2025. Afterwards, the final exam grades of the students will be introduced (2024 and 2025). Finally, the results of the crossover study will be presented and a preliminary hypothesis of possible factors and implications will be discussed.

This Section contains opinions directly quoted from the students. As such, the feedback received and incorporated in the various tables was not edited, with the exception of anonymizing the name of the AI assistant. Therefore, the reported students' opinions might include grammatical errors, typos, or syntactic inconsistencies. We chose not to correct these mistakes to preserve the originality of the feedback.

\subsection{What motivated students to use the LLM-enhanced assistant?}
In the experiments performed in the spring semester of 2024 and in the spring semester of 2025, students could voluntarily decide to perform the exercise with the chatbot or with the teacher. We administered a survey to students who chose the chatbot in the spring semester of 2025. In the survey, we asked the respondents to specify why they decided to volunteer to test the chatbot. The answers are presented in Table \ref{tab:motivation}.

\begin{table*}[]
\centering
\caption{Students' motivations for choosing the AI assistant. The chatbot's name was replaced by ChatGMP. The identified drivers are: curiosity (C), lower stress (LS), receive transcript (T), willingness to help (H), and others (O).}
\label{tab:motivation}
\begin{tabular}{|p{0.75\columnwidth}|c|} 
\hline
\textbf{Why did you decide to do the audit with the AI assistant?} & \textbf{Driver(s)} \\
\hline
A teacher said it was less anecdotal, easier to interpret. & O \\
\hline
We were curious of seeing how it would work with the ChatBot & C \\
\hline
The group decided so because we wanted to help in the ChatGMP development and also it was a WIN/WIN situation since we would also have an instructor present who would still be an auditee & H \\
\hline
To try out the technology, and have a more controlled environment for the audit. & C \\
\hline
I thought it would be a new experience to audit with ChatGMP than the regular ones in person. & C \\
\hline
Sounded innovative & C \\
\hline
The produced transcript would likely be easier to process than a verbal one. & T \\
\hline
It sounded interesting and I probably would've felt awkward essentially interviewing my professors. & C, LS \\
\hline
It seemed interesting to test it out. Moreover, it allowed us to take a bit of the pressure off compared to doing the audit with a lecturer/TA & C, LS \\
\hline
Wanted to try it out & C \\
\hline
Because I had heard about the project, and I know the course has a limited number of participants. I want to help remove this limit. & H \\
\hline
More comfortable asking ChatGMP than a teacher. & LS \\
\hline
Wanted to test something new. Also nervous to do the audit face to face with a teacher. & C, LS \\
\hline
Main reason was because it felt nice to have the whole transcript of the conversation, instead of having to listen through a recording to remember what was said during the audit. It was also a way of avoiding the awkwardness that probably would feel if I had to interrupt one of the teachers in the middle of an answer (that was not what we asked for). & T, LS \\
\hline
Because we feel it would be better to document the audit process and maybe less stress compare to audit with professors. & T, LS \\
\hline
To try something new & C \\
\hline
Our group wanted to help develop the chatbot so it can be implemented. We thought we would get a better experience since it doesn't elaborate as much as a real person. & H, C \\
\hline
\end{tabular}
\end{table*}

By making a theme analysis from the open answers displayed in Table \ref{tab:motivation}, the different drivers behind the choice to perform the exercise with the chatbot rather than with the physical teacher could be defined as: curiosity (C), lower stress (LS), wish to help the authors develop the initiative further (H), and obtaining a transcript of the conversation (T). \\
Most of the students (10 answers) express curiosity about the new technology and recognize that it is an innovative initiative. Some (6 answers) choose to interact with the chatbot because they felt less stressed than interviewing a teacher; other students also thought that they would get a better experience since a chatbot does not elaborate as much as a real person and they would not be put in the situation of having to interrupt a teacher asking to continue to the next question. Three students wanted to help us develop the model so that more students could enroll in the course the next years (currently, the limit is set to 120 students per year). Others (3 answers) thought that receiving the transcript of the interaction would be easier than having to take notes or listening to an audio recording. One answer presents other drivers (O), stating that the responses of the chatbot are less anecdotal and thus easier to interpret.

\subsection{Audit results: students' perspectives}

\subsubsection{Comparison between 2024 and 2025}

\begin{figure}[h]
\begin{multicols}{2}
    \subfloat[]{%
    \includegraphics[width=\linewidth]{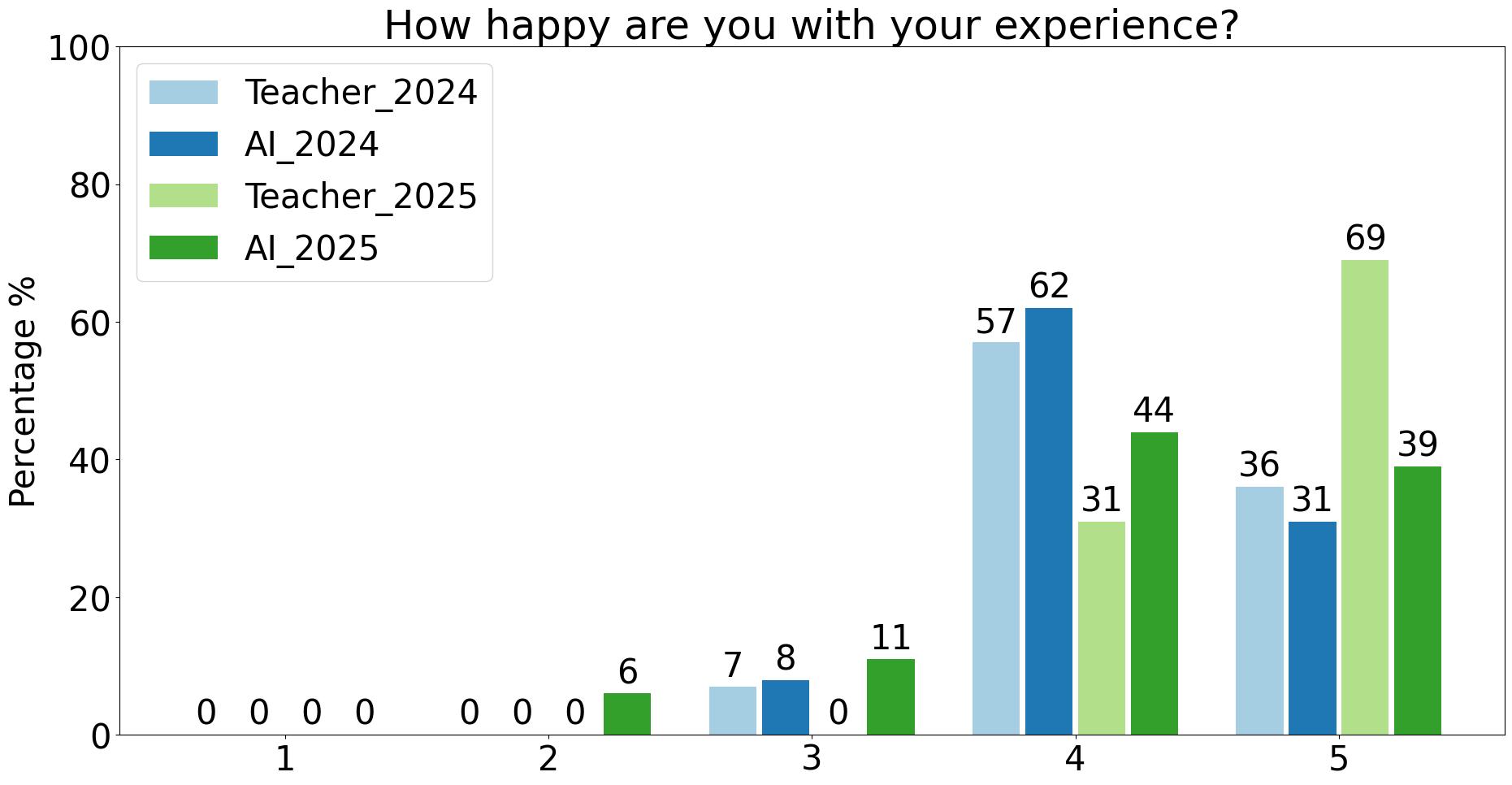}
    }\par
    \subfloat[]{%
    \includegraphics[width=\linewidth]{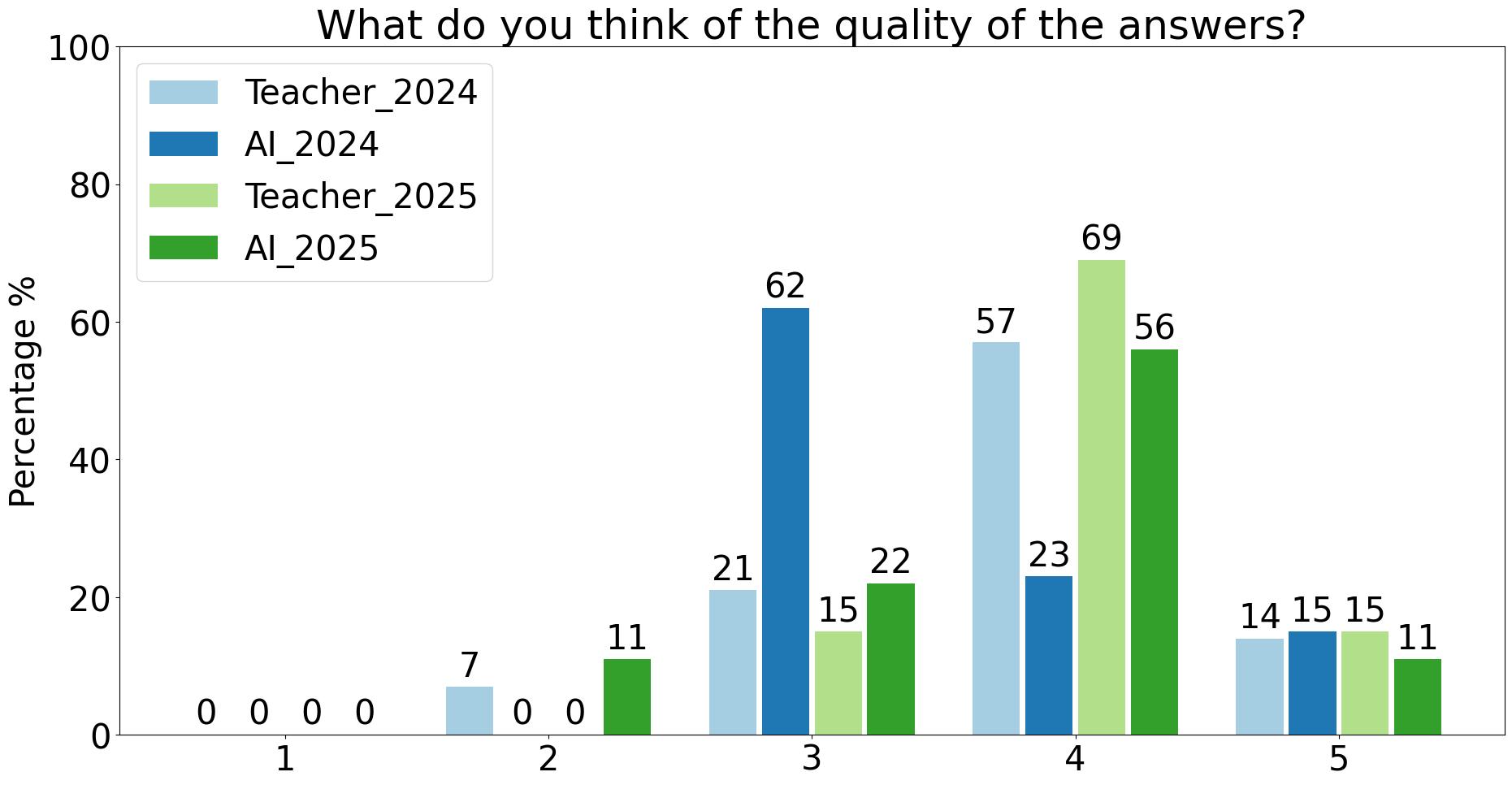}
    }\par
\end{multicols}
\begin{multicols}{2}
    \subfloat[]{%
    \includegraphics[width=\linewidth]{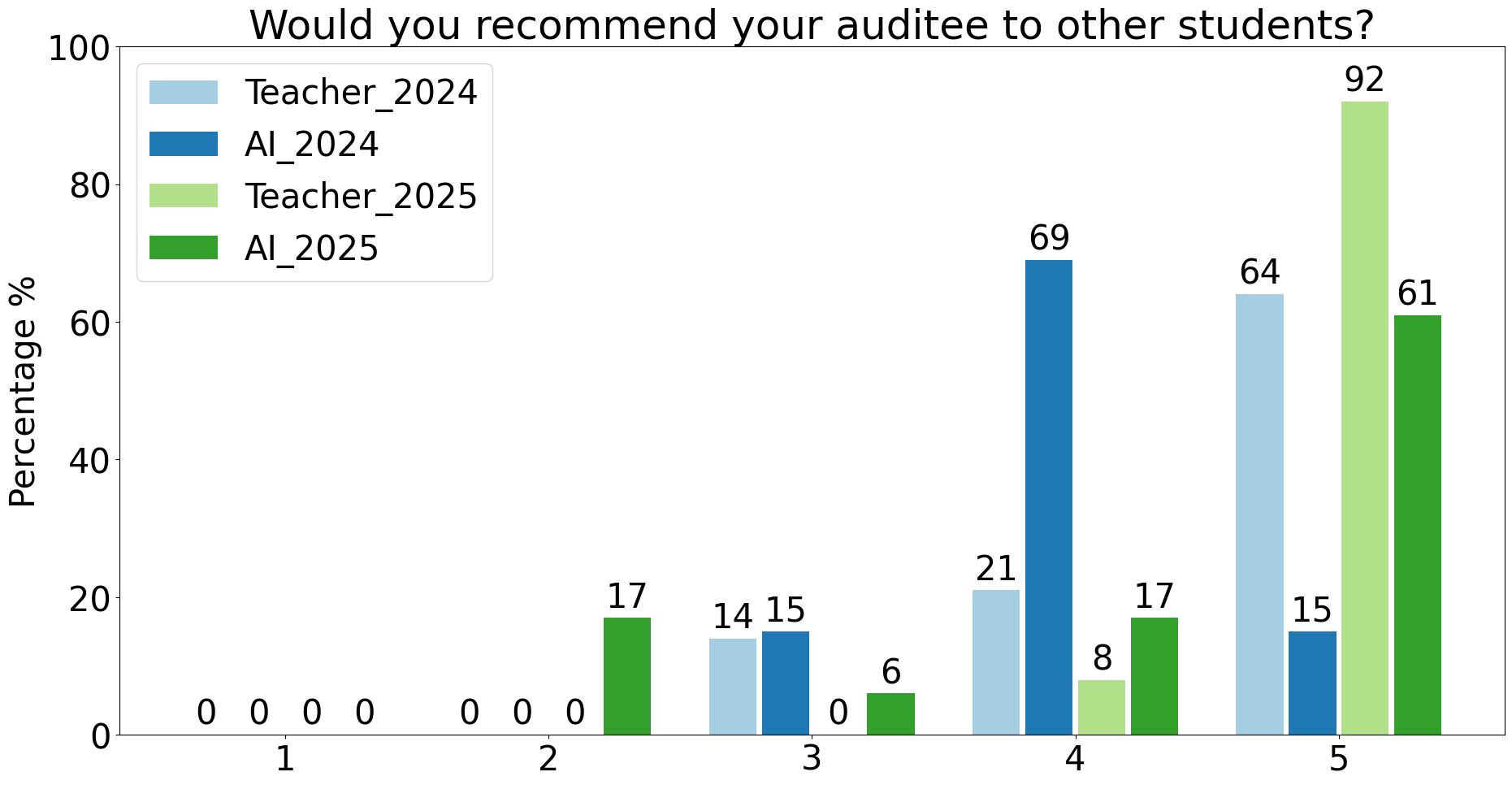}
    }\par
    \subfloat[]{%
    \includegraphics[width=\linewidth]{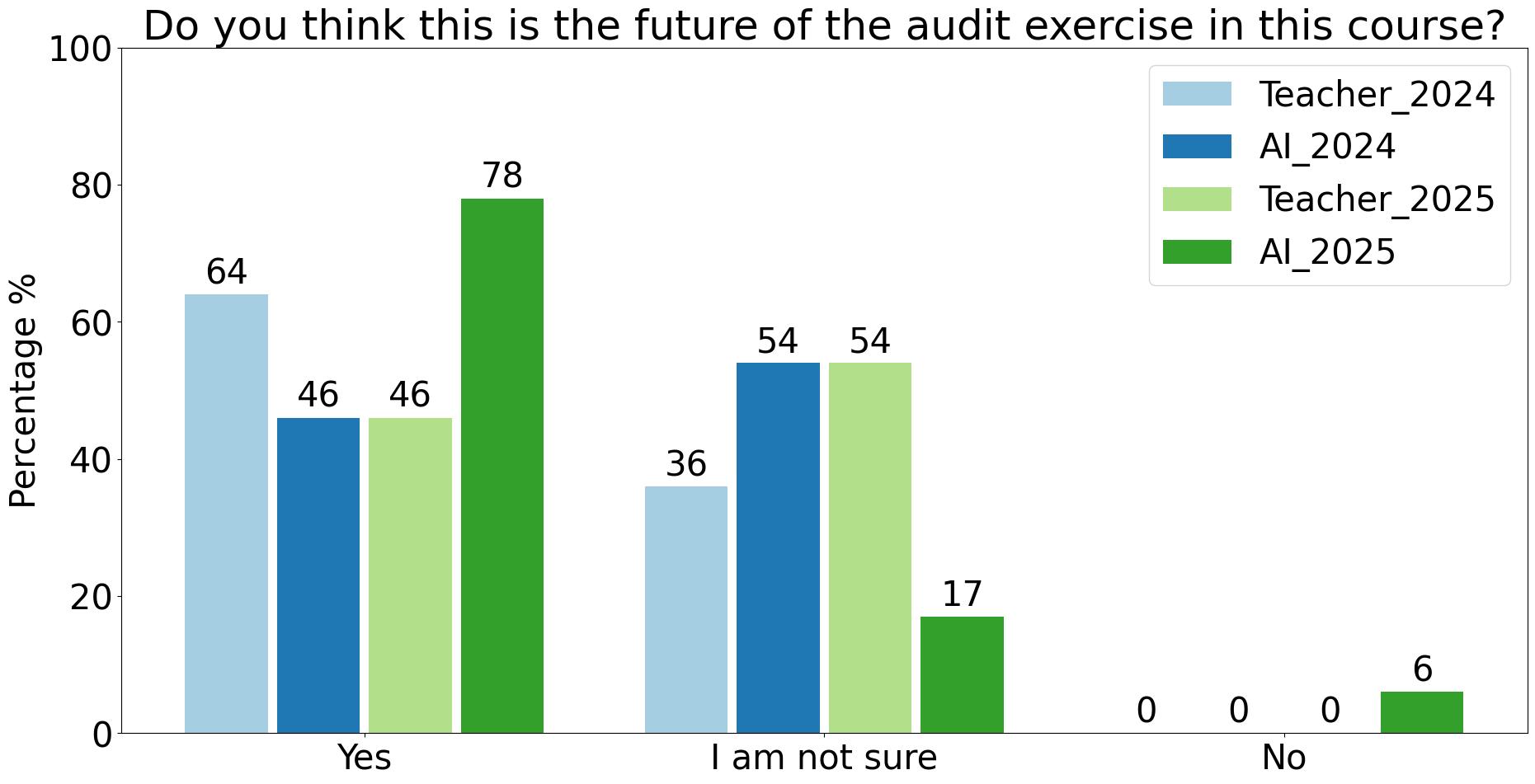}
    }\par
\end{multicols}
\caption{\label{fig:quest_students} Responses to surveys investigating the perceptions of students that performed the exercise with a teacher or with the LLM-enhanced assistant. Results are compared between 2024 and 2025. Responses to the research question (a) \textit{"How happy are you with your experience?"} (RQ1); (b) \textit{"After having performed the audit, what do you think of the quality of the answers?"} (RQ2); (c) \textit{"Would you recommend doing the audit with your auditee (teacher or chatbot) to other students?"} (RQ3); and, (d) \textit{"Do you think this is the future of the audit exercise in this course?"} (RQ4). RQ1, RQ2 and RQ3 are on a Likert scale in the range 1-5, from lowest to highest. One response is excluded because not relevant.}
\end{figure}

\begin{table*}[]
\centering
\caption{Mean and standard deviation of the groups auditing the teacher and the AI assistant for RQ1, RQ2, RQ3 and RQ4. The results are reported for the experiments conducted in the spring semesters of 2024 ($N_{teacher}=14$ vs. $N_{LLM}=13$) and 2025 ($N_{teacher}=13$ vs. $N_{LLM}=18$) and for the crossover study (CO) performed in the fall semester of 2025 ($N_{teacher}=6$ vs. $N_{LLM}=6$).}
\label{tab:mean}
\begin{tabular}{lcccc}
& \textbf{RQ1} & \textbf{RQ2} & \textbf{RQ3} & \textbf{RQ4} \\
& \textbf{avg.|std.} & \textbf{avg.|std.}  & \textbf{avg.|std.} & \textbf{avg.|std.} \\ \hline
\textbf{Teacher 2024} & 4.29; 0.59 & 3.79; 0.77 & 4.50; 0.73 & 4.29; 0.96 \\
\textbf{AI assistant 2024}  & 4.23; 0.58 & 3.54; 0.75 & 4.00; 0.55 & 3.92; 1.00 \\ \hline
\textbf{Teacher 2025} & 4.69; 0.46 & 4.00; 0.55 & 4.92; 0.27 & 3.92; 1.00\\ 
\textbf{AI assistant 2025}  & 4.17; 0.83 & 3.67; 0.82 & 4.22; 1.13 & 4.44; 0.77\\ \hline \hline
\textbf{Teacher 2025 CO} & 4.50; 0.50 & 4.33; 0.75 & 4.50; 0.75 & 4.67; 0.75\\ 
\textbf{AI assistant 2025 CO}  & 3.17; 1.07 & 2.33; 0.75 & 2.50; 0.96 & 3.00; 1.41\\ \hline
\end{tabular}
\end{table*}

\begin{table*}[]
\centering
\caption{Mann-Whitney U test of the distributions of the two samples for 2024 ($N_{teacher}=14$ vs. $N_{LLM}=13$) and 2025 ($N_{teacher}=13$ vs. $N_{LLM}=18$). We also report the results for the crossover study (CO) ($N_{teacher}=6$ vs. $N_{LLM}=6$). Mann-Whitney U test is performed for RQ1, RQ2 and RQ3 and Chi-square for RQ4. The differences are not statistically significant, expect for RQ3 in 2025, and RQ2 and RQ3 in 2025 CO.}
\label{tab:p_values}
\begin{tabular}{lcccc}
\hline
& \textbf{RQ1} & \textbf{RQ2}  & \textbf{RQ3} & \textbf{RQ4} \\ \hline
& \textbf{$U_{M}$|p-val} & \textbf{$U_{M}$|p-val}  & \textbf{$U_{M}$|p-val} & \textbf{Chi2|p-val} \\ \hline
\textbf{2024} & 95.5; .823 & 112.0; .284 & 123.0; .092 & 0.3; .576\\ \hline
\textbf{2025} & 158.5; .067 & 141.0; .281 & 175.5; .011* & 5.1; .077 \\ \hline \hline
\textbf{2025 CO} & 30.0; .055 & 34.5; .008** & 33.0; .014* & 4.5; .105 \\ \hline 
\end{tabular}
\end{table*}
The AI assistant was tested in the course with three groups volunteering for the first time in 2024. These preliminary results are presented in \cite{caccavale2025chatgmp}. 
Since then, more data was collected and the database of the model was enriched.
The results presented in Figure \ref{fig:quest_students} show that there is an improvement in students' perspectives regarding the AI assistant. The mean and standard deviation are summarized in Table \ref{tab:mean}. The Mann-Whitney U test \citep{mann1947test} results are presented in Table \ref{tab:p_values}.

For readability, and according to the year when the students were enrolled in the course, we will refer to the students that performed the exercise with a teacher as "Teacher\_2024" or "Teacher\_2025", and the students that performed the audit with the AI assistant as "AI\_2024" or "AI\_2025".

Regarding RQ1, there is an improvement, meaning that there are more students that report being very happy (score=5) in 2025, being 39\% compared to 31\% in 2024. However, it appears that more students compared to 2024 are also more unhappy (scoring 2 and 3 on the Likert scale).
The same trend is also observed for groups that performed the exercise with a teacher, where students reported being happier in 2025 compared to 2024.
Comparing the perspectives regarding the AI assistant in 2024 and in 2025 (second and fourth bars), there is a clear shift, where students appear to be more satisfied in 2025 about the quality of the answers (RQ2), if they would recommend the chatbot as auditee (RQ3) and if they think auditing the AI assistant is the future of the exercise (RQ4). 

Interestingly, students that interacted with the AI assistant in 2025 are very certain (78\%) that having an AI assistant is the future of the audit (RQ4). On the other hand, while 92\% of the "Teacher\_2025" group would very likely recommend their auditee (the teacher), compared to 61\% in the "AI\_2025" group (RQ3), they are less certain that auditing the teachers is the future of the exercise (RQ4). Table \ref{tab:p_values} presents the statistical tests performed to compare the distributions of the two groups (auditing the teacher and the chatbot). The only statistical significant result for the spring semesters is for RQ3 in 2025, where more students would highly recommend a teacher as auditor compared to the LLM.

The students were also asked to elaborate on their experience during the audit. Regarding RQ2, the integral feedback of the students who performed the audit with the AI assistant is presented in the Appendix, Table \ref{tab:feedback_RQ2}; the feedback of the students who performed it with the teachers is presented in Table \ref{tab:feedback_RQ2_teachers}. The "AI\_2025" groups highlighted that overall it was a very good experience and the chatbot could answer the majority of their questions. They also stated that the answers seemed "real" and made them feel like they were auditing a real company, helping them to become familiar with the audit process. Other students found prompting to be challenging since their questions had to be very specific. They also mentioned that sometimes the chatbot would provide repetitive answers.
The "Teacher\_2025" groups also found the audit experience to be good and realistic overall. However, they noticed that the teachers sometimes spoke too extensively or answered unasked questions to gain time, and they felt it was rude to interrupt them. They also felt some of the information requested was not provided (since this is part of their learning).

\begin{table}[]
\centering
\caption{Students' perspectives regarding the future of the auditing exercise.}
\label{tab:feedback_RQ4}
\begin{tabular}{|p{0.95\columnwidth}|} 
\hline
\textbf{RQ4: Do you think this is the future of the audit exercise in this course?} \\
\hline
If it increases the throughput then it would be the future, i do not know if i am qualified to answer this as i did not try a audit with a teacher. i think you lack the human part. in some way it felt like just asking for documents. \\
\hline
I see the potential in using a chatbot to be able to work with more students, but I believe a lot of work needs to be done for it to feel close to a real audit \\
\hline
why, because more and more students will be able to take this very important course \\
\hline
It has a lot of potential, it saves time and it works pretty good. If it gets more training, it will be even better. \\
\hline
I think it is a good alternative and advanced option for the process of auditing. \\
\hline
If the chat gets improved, it could definitely be the future, but at this point then no. \\
\hline
Innovative \\
\hline
The pure efficiency and accuracy that comes with increased data - means that future audit exercises will continue to produce better and better results. \\
\hline
I think having the option to do either is advantageous, though it obviously requires more manpower. \\
\hline
I think it works fine and will open the course up to more students in the future. Also for its purpose as an exercise rather than an exam. \\
\hline
It is still missing the personal interaction part, that you would meet in a real audit. \\
\hline
This is the necessary future, if the course should be free of any participation limits, and I think it is well on the way. It was nice having the teaching assistant there, but not necessary at all. \\
\hline
A teacher may not always be available and reduces teacher’s workload \\
\hline
More people can take part in the course if this is used. This could also be something open to students outside of this course in order to test and learn more. \\
\hline
With a little work I really think this could be a great way to do all the audits in the future. \\
\hline
It’s good. \\
\hline
I don't think it's efficient \\
\hline
I think it would be really nice that everybody could take the course and people understand why they would have to use the chatbot. \\
\hline
\end{tabular}
\end{table}

Table \ref{tab:feedback_RQ4} presents the opinions of the "AI\_2025" groups. For comparison, the perspectives of the "Teacher\_2025" groups are presented in the Appendix, in Table \ref{tab:future_teacher}. The students think that the chatbot solution we developed is innovative and has the potential to become the future of the exercise, since this would allow more students to enroll in the course while maintaining the same level of educational quality and rigor. On the other hand, some students point out that the current chatbot version has some limitations that should be addressed in order to be able to fulfill its purpose. Additional feedback is included in Table \ref{tab:feedback_additional_info}.

\subsubsection{Advantages and disadvantages of the LLM-enhanced assistant}

Summarizing the feedback presented in the surveys (Tables  \ref{tab:feedback_RQ4}, \ref{tab:feedback_RQ2}, \ref{tab:feedback_additional_info}) together with the opinions expressed by the students in the audit reports, we derive two main trends: (i) advantages of using the AI assistant; and, (ii) issues related to this specific model. These points are summarized below.

\vspace{0.5cm}
\noindent\textbf{Advantages according to students:}\\
\begin{itemize}
    \item The initiative is highly innovative, and many students cited "curiosity" as the driver behind their choice to interact with the AI assistant rather than with a teacher.
    \item All students should have the possibility to enroll in the course, while now there is an uptake limit of 120 people; therefore, the AI assistant could pose a valuable solution and minimize the impact of this limitation. 
    \item Interacting with an LLM reduces the pressure and anxiety typically associated with preparing and performing the exercise for the first time. They might feel less pressured to ask questions to an AI assistant rather than a teacher.
    \item Receiving the full transcript of the conversation. While the teams that audit the chatbot receive a transcript of the conversation, the other groups receive an audio recording, which they might find more cumbersome. 
    \item The chatbot allowed students to use the time between questions to discuss within the team without pressure. During the audit, for each question, the students had time to review the answer and discuss in the group how to proceed. This might spur collaborative learning and team effort. 
    \item The decreased pressure and stress also encourages to read the answers before asking the next question, while in the audit with the teacher, the students tend to rush to the next question because they rely on the recording.
\end{itemize}
Given these advantages, they concluded that the LLM-based chatbot has the potential to become a very useful tool and can help save time.
On the other hand, students also point out various problems, issues and challenges connected with the LLM-enhanced assistant.

\vspace{0.5cm}
\noindent\textbf{Disadvantages and identified issues:}\\
\begin{itemize}
    \item Questions have to be phrased as simple as possible, and that the LLM is not at the same level at comprehending complex or related questions, compared to the teacher, i.e., it does not understand the similarity between words such as "document" and "procedure". This limitation is given by the relatively small size of the LLM employed, as correctness was prioritized to generative power when developing the tool. 
    \item Since the model bases its knowledge on historical knowledge, it is sometimes unable to answer niche or unseen questions.
    \item The model should be more clear about the certainty of the information given instead of providing (factually correct but) unrelated answers. 
    \item Sometimes the model generated identical answers to slightly different questions and repeated the answers, providing the same information, even when the students rephrased the questions. 
    \item A commonly cited issue is that the AI assistant did not remember the whole conversation, so it was cumbersome for students to have to specify the context again for follow-ups.
    \item Another issue being raised was the fact that the students could not "copy and paste" the questions. 
\end{itemize}

Given these disadvantages and issues, many students concluded that, although the tool has potential, it should be improved before it is properly effective.

Furthermore, students also provided some ideas for potential improvements, such as (i) to include a virtual tour and images of the facility where the operations take place, and (ii) to include voice recognition, so that the AI assistant could process oral questions (speech-to-text model) and read aloud the answers (text-to-speech model).

\subsection{Students' performance}

\begin{table*}[]
\centering
\caption{Average grade of the students performing the audit exercise with the AI assistant and the ones performing the audit with the teacher. Grades are reported according to the University's grading system, where the minimum passing grade is 2 and the maximum grade is 12.}
\label{tab:performance}
\begin{tabular}{lcccc}
\textbf{Year} & \textbf{Avg. with AI assistant} & \textbf{Avg. total}  \\ \hline

\textbf{2024} & 6.8 & 7.9 \\ 
\textbf{2025}  & 8.7 & 9.1 \\ \hline
     
\end{tabular}
\end{table*}

Table \ref{tab:performance} presents the average final grade for the students performing the exercise with the AI assistant. 
The marks follow the Danish grading system \citep{danish_grades} and are publicly available \citep{gmp_grades}. 
Of note is that, while in 2024 the report students handed in after interacting with the LLM was graded and accounted for 25\% of the grade, in 2025 it was only Pass/Fail, therefore the grade is disentangled from the report and thus the interaction with the LLM. 

The results show a slightly lower grade average for the students who interacted with the AI assistant, compared to the total average. Generally, we see that students belonging to the same groups (who did the audit and wrote the report together) tend to receive the same final grade. However, this is not a strict indication of the grades, since one group presented a student with the maximum grade and one with the minimum. 
This may suggest that the AI assistant provided all the necessary information for the groups to succeed in the learning objectives of the exercise and in writing a good report; however, the final grade is up for individual study. We do not have other grade metrics that would allow us to characterize the students further, such as an overall average grading for the entire education. Therefore, this might be a pattern that highlights the type of students who choose to perform the audit with the chatbot (e.g., perhaps students that aim to the top grade would tend to select the teacher rather than testing a new tool). However, this hypothesis needs further investigation and no conclusions can be drawn at this stage, neither whether the chatbot led to lower grades nor that higher-achieving students tend to choose to audit the teacher.

\subsection{Incorporating students' feedback: improving the GUI}
\label{incorporating_feedback}

\begin{figure*}[]
\centering
\includegraphics[width=1\linewidth]{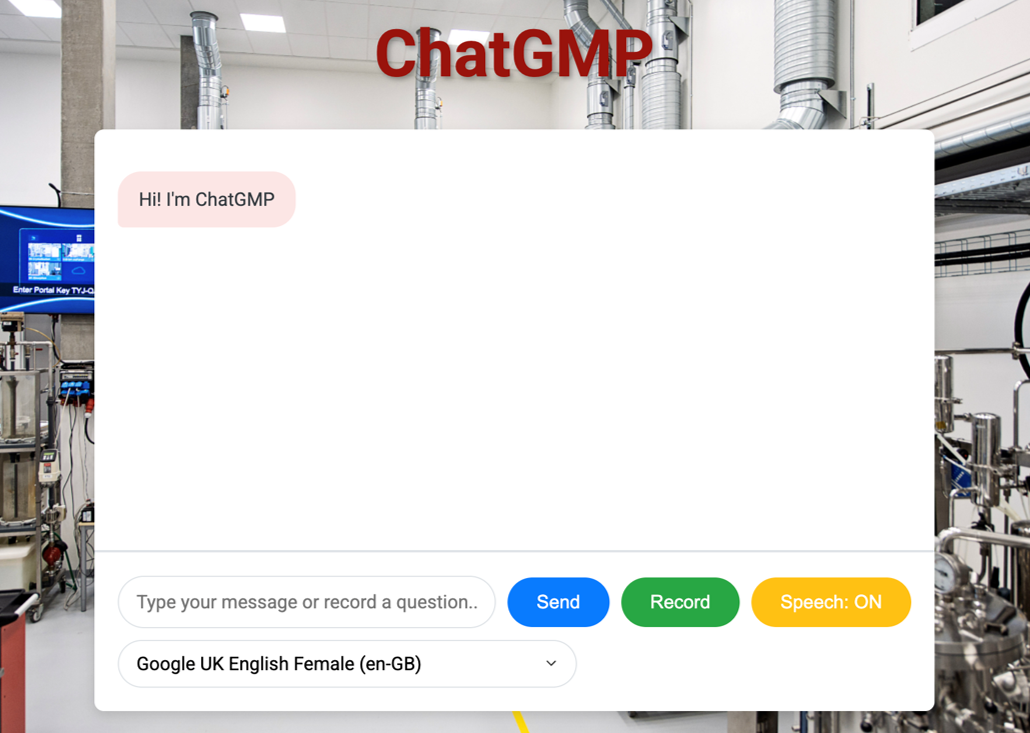}
\caption{\label{fig:new_chatgmp}New graphical interface of the AI assistant.}
\end{figure*}

Following the feedback received in the spring semester of 2025, to improve the general experience of the students, a few modifications and additional functionalities were implemented: (a) improved GUI of the model, as presented in Figure \ref{fig:new_chatgmp}; (b) embedded a speech-to-text function, so that students could record their questions to the AI assistant, instead of typing; (c) embedded a text-to-speech function, so that the response generated by the chatbot would be read out (users can decide whether to turn off this option, by pressing the "Speech: OFF" button, and which voice they prefer to hear, in case the speech option is on.); and, (d) allowed students to "copy and paste" their questions, for time saving to avoid penalizing the slow-typing students. 

\subsection{Results of the crossover study}

\begin{figure}[h]
\begin{multicols}{2}
    \subfloat[]{%
    \includegraphics[width=\linewidth]{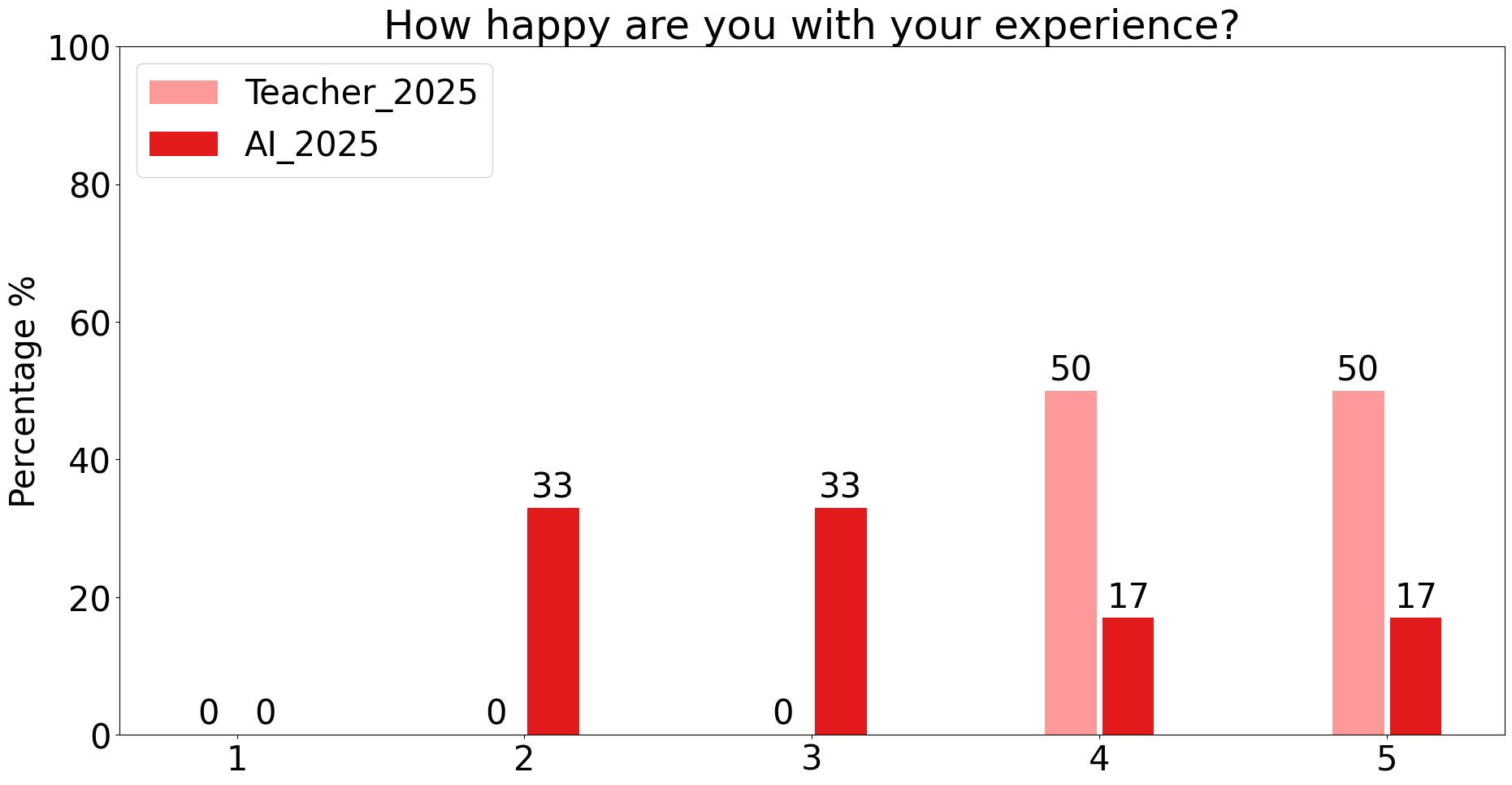}
    }\par
    \subfloat[]{%
    \includegraphics[width=\linewidth]{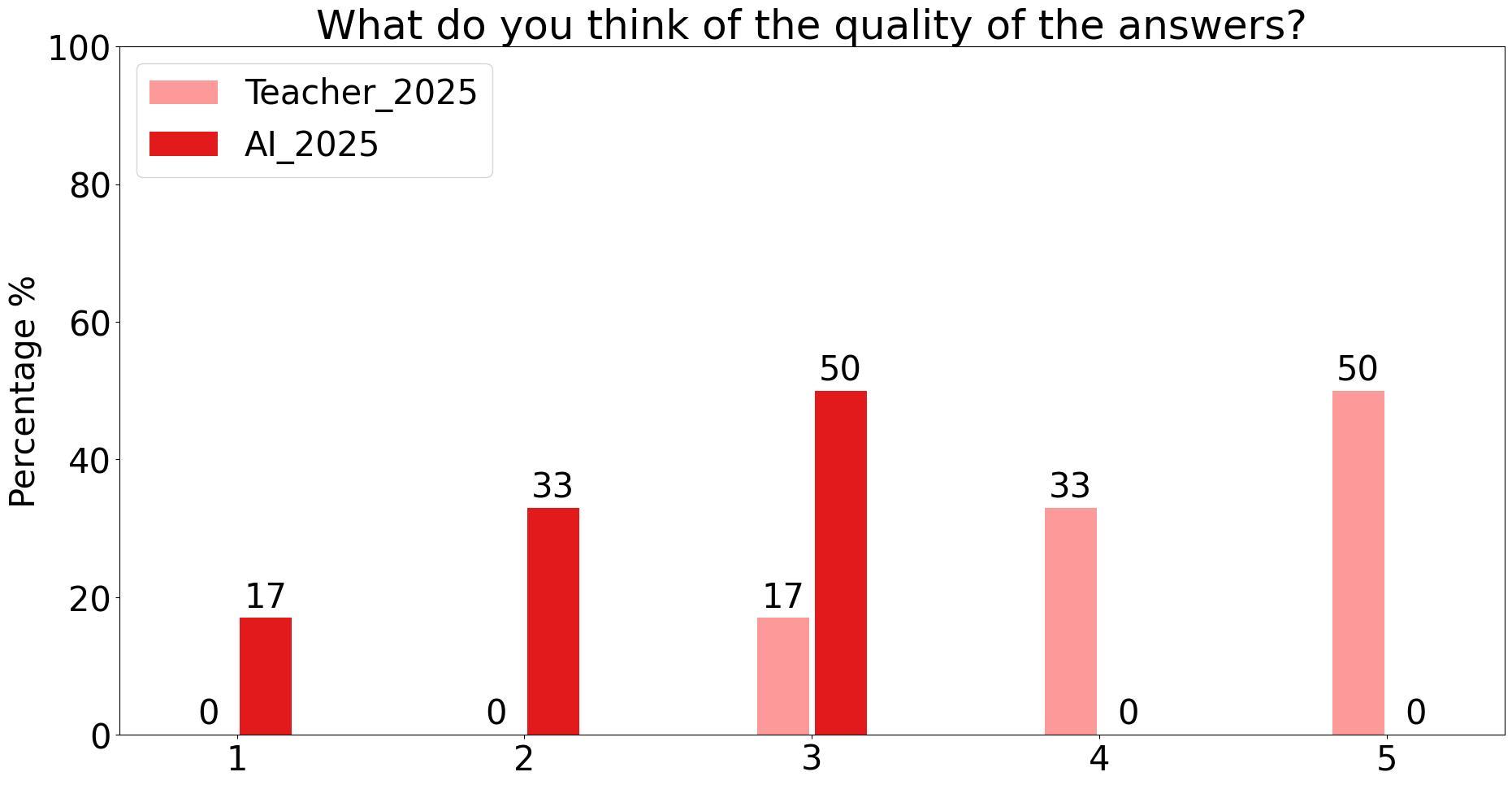}
    }\par
\end{multicols}
\begin{multicols}{2}
    \subfloat[]{%
    \includegraphics[width=\linewidth]{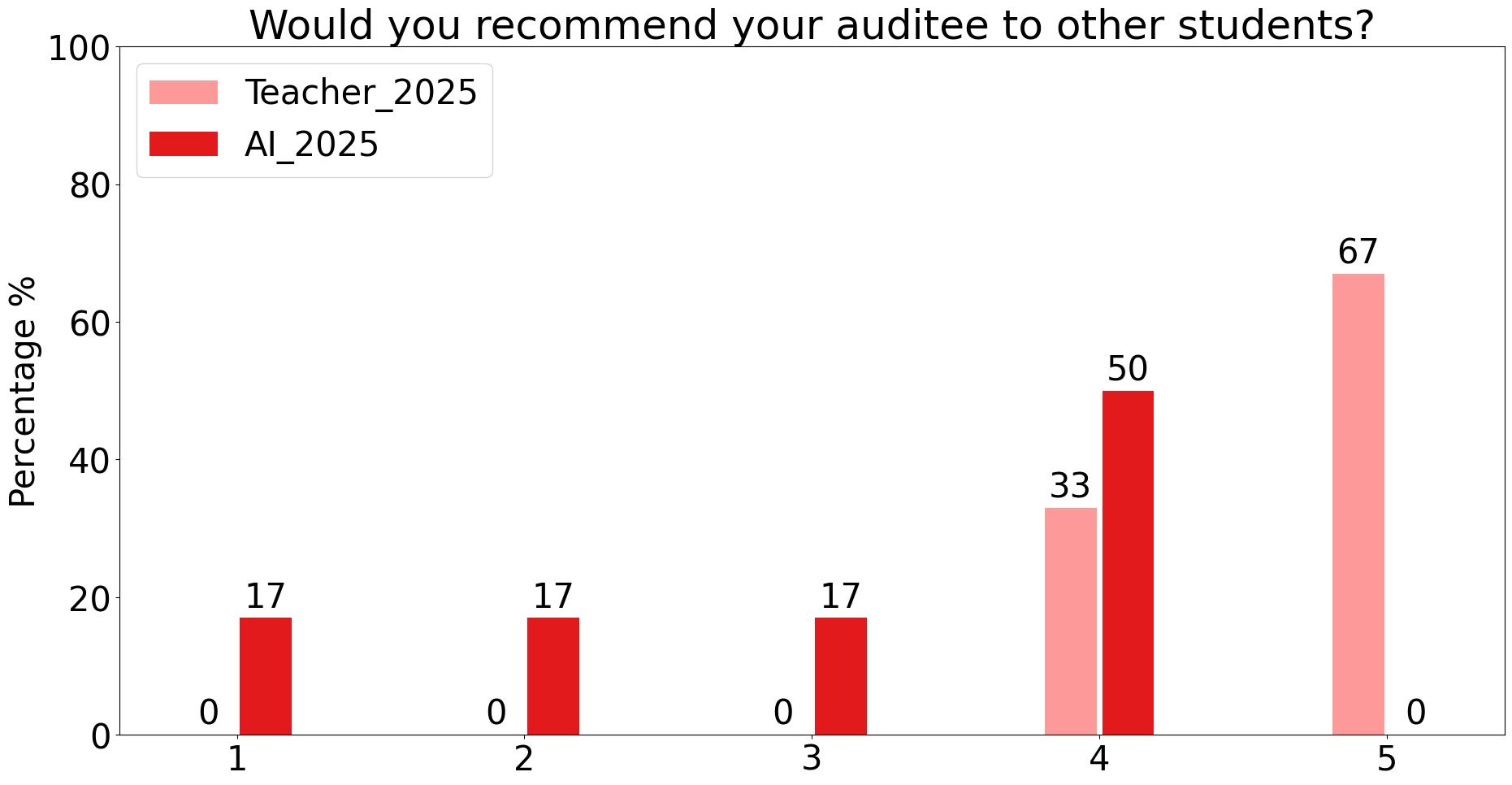}
    }\par
    \subfloat[]{%
    \includegraphics[width=\linewidth]{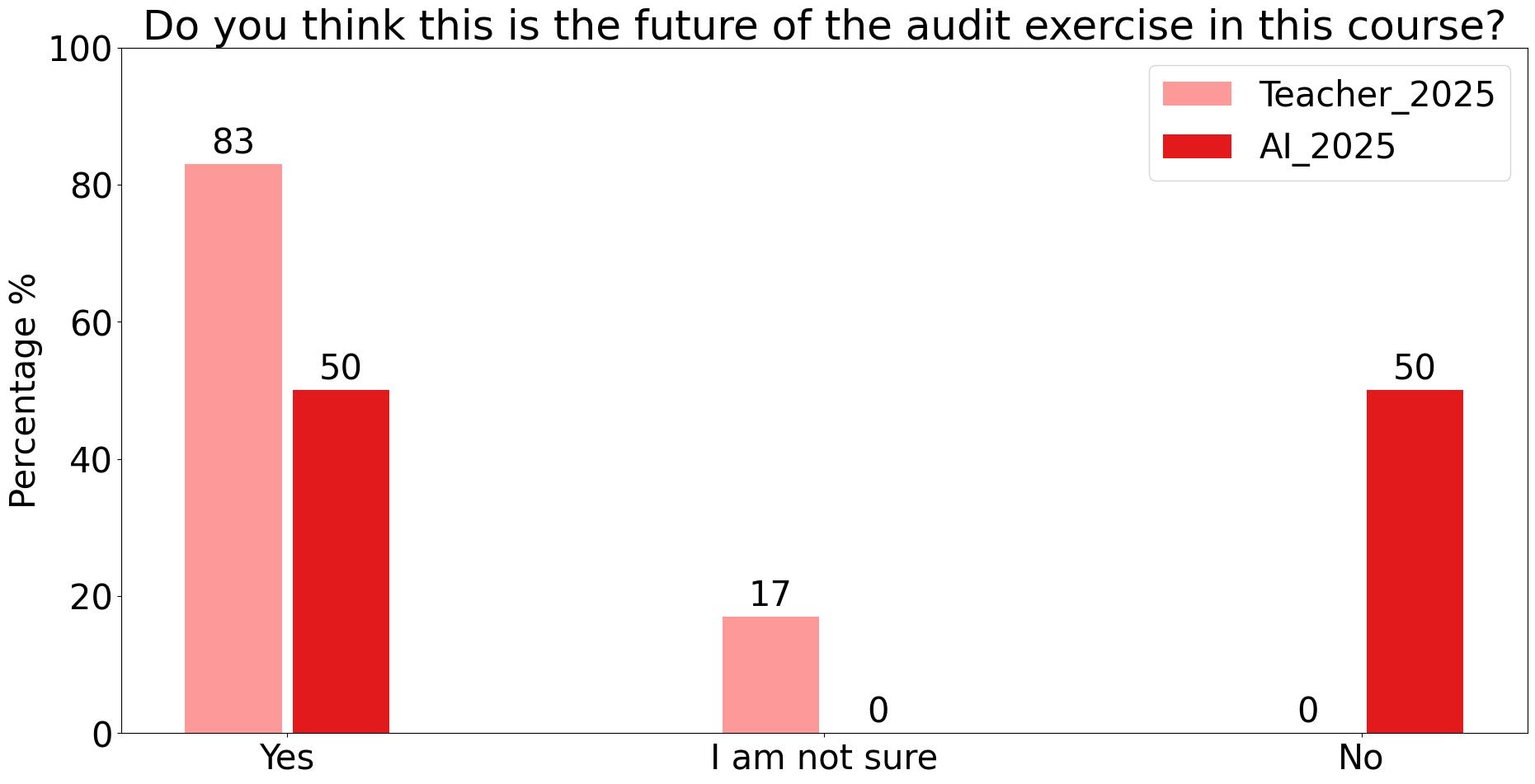}
    }\par
\end{multicols}
\caption{\label{fig:results_fall2025} Responses to surveys investigating the perceptions of students that performed the exercise with a teacher or with the LLM-enhanced assistant. Results are reported for the crossover study performed with six students, divided in two groups of three students each, in the fall semester of 2025.}
\end{figure}

After incorporating some of the students' feedback received from the previous iteration of experiments, we tested the model again in a crossover study. 

The results of this experiment, conducted in the fall semester of 2025, are presented in Figure \ref{fig:results_fall2025}, as well as summarized in Table \ref{tab:mean}. The results appear to be quite different from those obtained during the spring semesters of 2024 and 2025. Here, in fact, students seem to prefer the interaction with the teacher, finding the quality of the responses (RQ2) to be quite higher (on average 4.50 compared to 3.17 with the AI assistant) and are less sure that the chatbot can effectively replace the teacher (RQ4) moving forward (4.67 compared to 3.00). Table \ref{tab:p_values} shows the results of the statistical tests performed. In the crossover study, results are statistically significant for RQ2 (p-value < 0.01) and RQ3 (p-value < 0.05).

\begin{table}[]
\centering
\caption{Perceptions of the students regarding the different auditees.}
\label{tab:preference}
\begin{tabular}{p{0.23\columnwidth}|p{0.15\columnwidth}|p{0.15\columnwidth}|p{0.15\columnwidth}|p{0.15\columnwidth}}
\textbf{Question} & \textbf{Prefer the teacher [\%]} & \textbf{Prefer the chatbot [\%]} & \textbf{Both were good [\%]} & \textbf{Neither was good [\%]} \\ \hline
\textbf{Did you prefer the audit with the teacher or chatbot?}  & 66.7 & 0 & 33.3 & 0 \\ \hline
     
\end{tabular}
\end{table}

Overall, Table \ref{tab:preference} shows that 66.7\% of students preferred the audit with the teacher, while 33.3\% stated that both were good. Further opinions clarifying their reflections are provided in 
Table \ref{tab:feedback_crossover}. Here, students find the answers given by the ChatGMP to be useful in finding topics and a good tool to use for students who just want a quick and easy non confrontational way to get answers. However, they recognize that the audit with the teacher was quite close to a real life audit. 

Table \ref{tab:good_bad} presents a comparison of the topics in which each auditee was particularly good or bad. Students find the teacher to be better at providing rich and more nuanced answers (and at "improvising"), although they find this setting to be quite time-intensive and, as a result, they do not manage to exhaust their list of questions. On the other hand, the chatbot is good at providing documents and fast answers, but sometimes struggles to understand the questions if they are complex or unseen.

Table \ref{tab:differences_teacher_chat} presents a further comparison between the teacher and chatbot. One of the students states that the "two auditees together actually complement each other well, as the [teachers] provide context, accountability, and access to real records, while the ChatGMP auditee accelerates scoping, and reaching requested
documents". This is also supported by another opinion reported in Table \ref{tab:feedback_crossover}, where a student suggested that, given the different nature of the experience, both audit experiences should be offered, since many other students might prefer the AI interaction.

\begin{table}[]
\centering
\caption{Students' reflections in the crossover study.}
\label{tab:feedback_crossover}
\begin{tabular}{|c|p{0.45\columnwidth}|p{0.45\columnwidth}|} 
\hline
& \textbf{Teacher} & \textbf{AI assistant} \\
\hline
\textbf{RQ2} &
Some answer where good, some answers a bit too theoretical. We could benefit from a batch report being available as documentation.
\vspace{0.5em}
\newline
The answers were meant to guide us away from missing or inadequate information, which could be realistic, for our team it was difficult to cut off the answers and go back on track.
&
The ChatGMP answers was useful to locate topics, but most responses were generic and not suitable as primary evidence. In other words, the answers were informative but incomplete/sufficient to flag risks, not sufficient to demonstrate robust, effective GMP control without follow-up documentation and validation.
\vspace{0.5em}
\newline
It can clearly be seen that there is a big difference between human interaction and AI. For example there were cases that if asked to elaborate on a question it would get confused or it would not understand a question asked.
\\
\hline
\textbf{RQ4} &
Teacher is better at giving longer answers and explaining the documents. This is also more fun than ChatGMP. Gives a more authentic audit experience. I think it should be offered along with ChatGMP for those that want it. I'm sure you will get many that prefer ChatGMP.
\vspace{0.5em}
\newline
Compared to Chat, the teacher gave more interesting / elaborate answers, but considering the workload, ChatGMP is a better option to open up the course to others
\vspace{0.5em}
\newline
Overall the audit was quite close to a real life audit and I think that for students that do not know what an audit is it is a very nice introduction to the topic.
&
The AI assistant is a good tool to use for students who don't need the physical experience or just want a quick and easy non confrontational way of getting answers. Using ChatGMP in a time limited situation, makes us just fire off all our questions quickly not reading the responses or asking deeper follow up questions.
\vspace{0.5em}
\newline
It is able to generate fine answers based on the transcript received from previous audits, but it probably needs some more diverse knowledge / more transcripts to learn from.
\vspace{0.5em}
\newline
As it is a nice idea to have an AI, I do not believe that this is the future of an audit as it requires human interaction. There are a lot of topics that an AI could not answer or understand and in general it would be difficult to implement real life as some questions might be unexpected and the auditors will not receive the answers they are looking for.
\\
\hline
\end{tabular}
\end{table}

\begin{table}[]
\centering
\caption{Students' feedback on topics where the auditee (teacher or AI assistant) was either particularly good or bad at handling.}
\label{tab:good_bad}
\begin{tabular}{|p{0.1\columnwidth}|p{0.44\columnwidth}|p{0.44\columnwidth}|} 
\hline
& \textbf{Teacher} & \textbf{AI assistant} \\
\hline
\textbf{Any topic that was particularly good at?} &
GMP rules
\vspace{0.5em}
\newline
Rich context and nuance; there were possibility to read body language, push back, and pivot in real time. Beside quick access to primary records (screenshares, binders) and on-the-spot walk-through. Presenting mix of procedures [...], records [...], and Management Review made the exercise feel realistic, and all documents examples were good.
\vspace{0.5em}
\newline
Overall the audit interview was really nice and had a nice flow as well as presenting the documentations.
&
Giving documents.
\vspace{0.5em}
\newline
was relatively fast and consistent was good for scoping, checklists, and clause mapping. good in term of reduce note-taking load. easy to locate topics and draft follow-ups. No scheduling friction
\vspace{0.5em}
\newline
AI was very good in regards to documentation and I believe that that could be a great focus area if it is planned to be used in companies. It was very nice to receive the documents so fast so a database with all company documents could be built in so when inspectors come they can get any documents they wish on the spot.
\\
\hline
\textbf{Any topic that was particularly bad at?} &
Process knowledge
\vspace{0.5em}
\newline
Time-intensive; answers may be long. [...] i think the weakest part was evidencing: too many answers were narrative without primary records, some documents were low-resolution, which made clause-level verification hard. The discussion basically leaned on narrative discussion without first modeling to what “good evidence” and a defendable finding look like. the discussion missed a couple of “checkpoints” where i think teacher can give introduction before interview to good interview technique.
\vspace{0.5em}
\newline
Not all information was received stating that it is confidential [...]
but in real life I am nor aware that we could say to an inspector that we cannot give a certain information as they would consider that the company is hiding something from them causing more problems.
&
Understanding things for which it was not specifically programmed.
\vspace{0.5em}
\newline
Responses sometimes were generic
\vspace{0.5em}
\newline
there were a lot of questions unanswered or AI did not understand the question.
\\
\hline
\end{tabular}
\end{table}

\begin{table}[]
\centering
\caption{Students' feedback on the main differences between performing the audit with the teacher and with the chatbot.}
\label{tab:differences_teacher_chat}
\begin{tabular}{|p{0.95\columnwidth}|} 
\hline
\textbf{Can you elaborate on the main differences between performing the audit with the teacher and the chatbot?} \\
\hline
flow, time consumption from ChatGMP without results \\
\hline
The teacher part was more fun and was heavier on role play. Easier to find weaknesses in the answers and process. \\
\hline
With ChatGMP we had better pace as it had fix answers that we just skimmed through and either followed up or moved on. The teacher gave long, more elaborate stories, but it was longer answers therefore we were able to ask less questions. \\
\hline
I think using the two auditees together actually complement each other well, as the managers provide context, accountability, and access to real records, while the ChatGMP auditee accelerates scoping, and reaching requested documents. \\
\hline
Easier flow during the interview, more elaborate and clear answers and the answers were understood by the teacher while not all of them were understood by AI. \\
\hline
there was no flow and the feeling of real audit in the chatbot \\
\hline
\end{tabular}
\end{table}

\begin{figure*}[]
\centering
\includegraphics[width=0.8\linewidth]{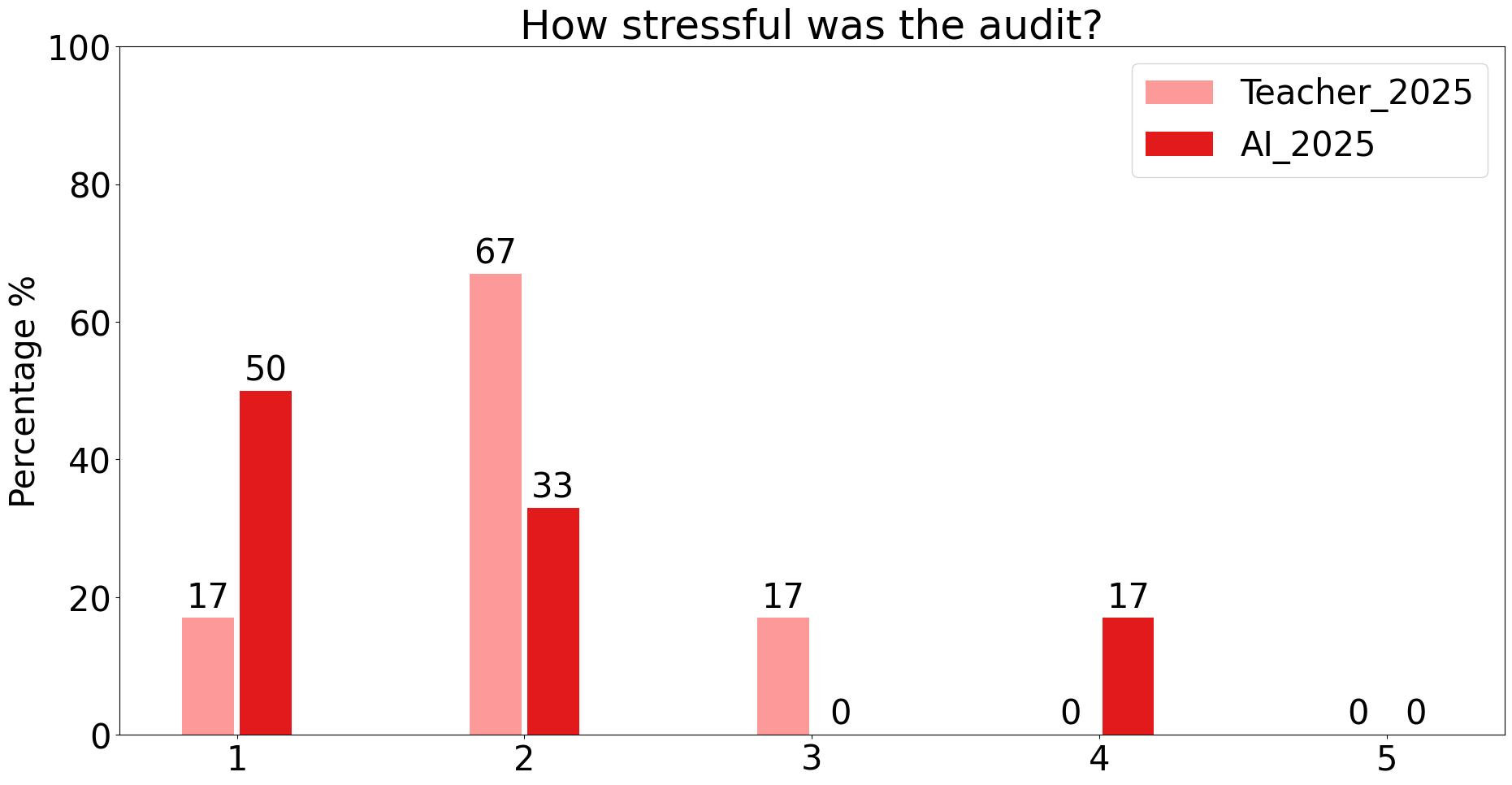}
\caption{\label{fig:stress}Stress perception of the students when performing the audit with the teacher and AI assistant as auditees. Results are on a Likert scale between 1 (not stressful at all) and 5 (very stressful). Average stress with the teacher is 2, with the AI assistant is 1.8 (p-value=0.485).}
\end{figure*}

Validating some of the feedback collected in 2024 and 2025, we wanted to test whether students found the audit with the AI assistant less stressful than with a teacher. Figure \ref{fig:stress} shows that most students find the audit with the teacher to be more stressful, with an average of 2.00 (standard deviation 0.58) on the Likert scale, compared to 1.83 (standard deviation 1.07) with the AI assistant. However, results are not statistically significant (p-value=0.485).

\begin{figure*}[]
\centering
\includegraphics[width=0.8\linewidth]{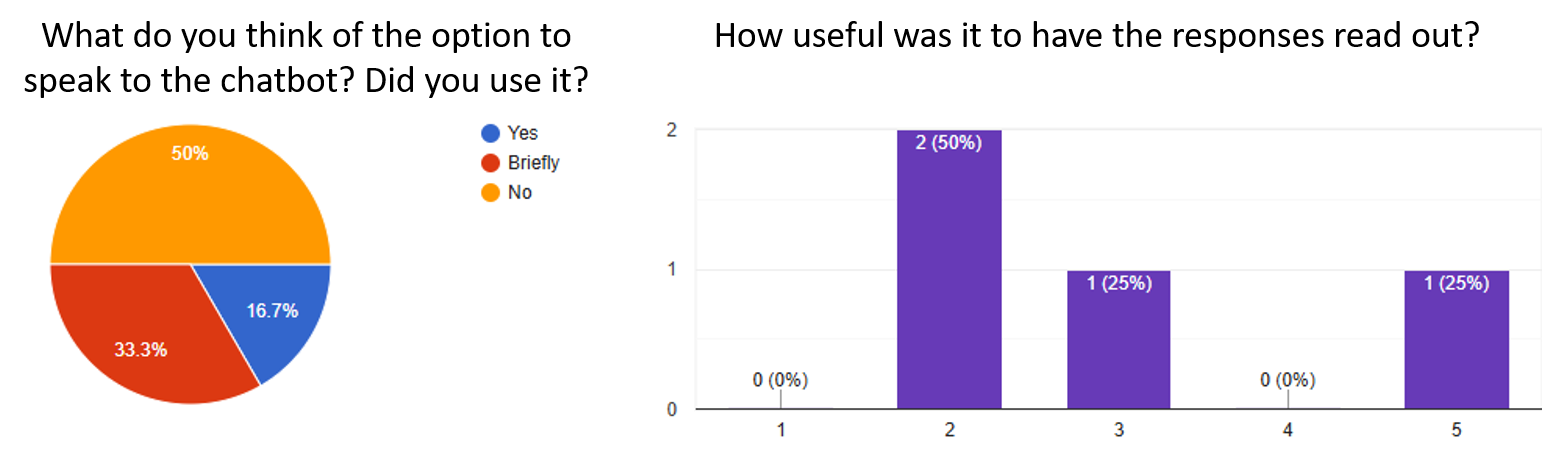}
\caption{\label{fig:speak}Students' perception on the speech-to-text and text-to-speech implementations in the chatbot. Results of the second graph are presented on a Likert scale from 1 (not useful at all) to 5 (very useful).}
\end{figure*}

Finally, as mentioned in Section \ref{incorporating_feedback} based on the provided students' feedback, we have also integrated text-to-speech and speech-to-text into the AI assistant. We asked students whether they tried the functions and if they were useful, the results are illustrated in Figure \ref{fig:speak}. It seems that the students preferred to type rather than speak to the AI assistant, thus not finding the function useful. We argue that this might be due to the fact that they had prepared their questions in advance and could copy and paste them.

\subsection{Teachers' feedback}
In addition to performing audits with students, the tool could also be used to train new teachers who join the course and need to prepare for the audits. The teacher was given access to the chatbot, so that they could train with the help of the AI assistant. 
The teacher provided the following feedback: 
"I enjoy the tool a lot, and it helped me initially to verify if my answers make sense. The only downside was the waiting time, which wasn't too bad. I also liked how easy it was to install it on my computer! I see it as the future of the audit exercise and would like to see it developing in the future (with videos etc.)". 

\section{Discussion}
\label{discussion}

This section aims to contextualize the results presented in this study in light of the general landscape of AI in education, encompassing the opportunities, challenges, and future directions that LLM-enhanced teaching and learning could bring.

\subsection{AI assisted collaborative learning: takeaways of the study}

This study focuses mainly on LLM-enhanced collaborative learning, discussing a custom AI assistant developed for an M.Sc. course. The results highlight various trends regarding the use of LLMs in Higher Education.  

Firstly, among the aspects that motivated the students to use AI was curiosity: they thought the initiative was interesting and innovative, and they wanted to test the developed tool first-hand. They found the interaction with an LLM to be less stressful than interviewing a teacher. They also wanted to help us with the development and evaluation of the tool, since this tool would remove the limitation of yearly students uptake for the course. Finally, receiving a written transcript of their conversation rather than an audio recording was also one of the main drivers. Although these motivations are intrinsic to this specific application and cannot necessarily be separated from the course, they do suggest some general trends that may also apply to other domains.

\begin{figure*}[]
\centering
\includegraphics[width=0.8\linewidth]{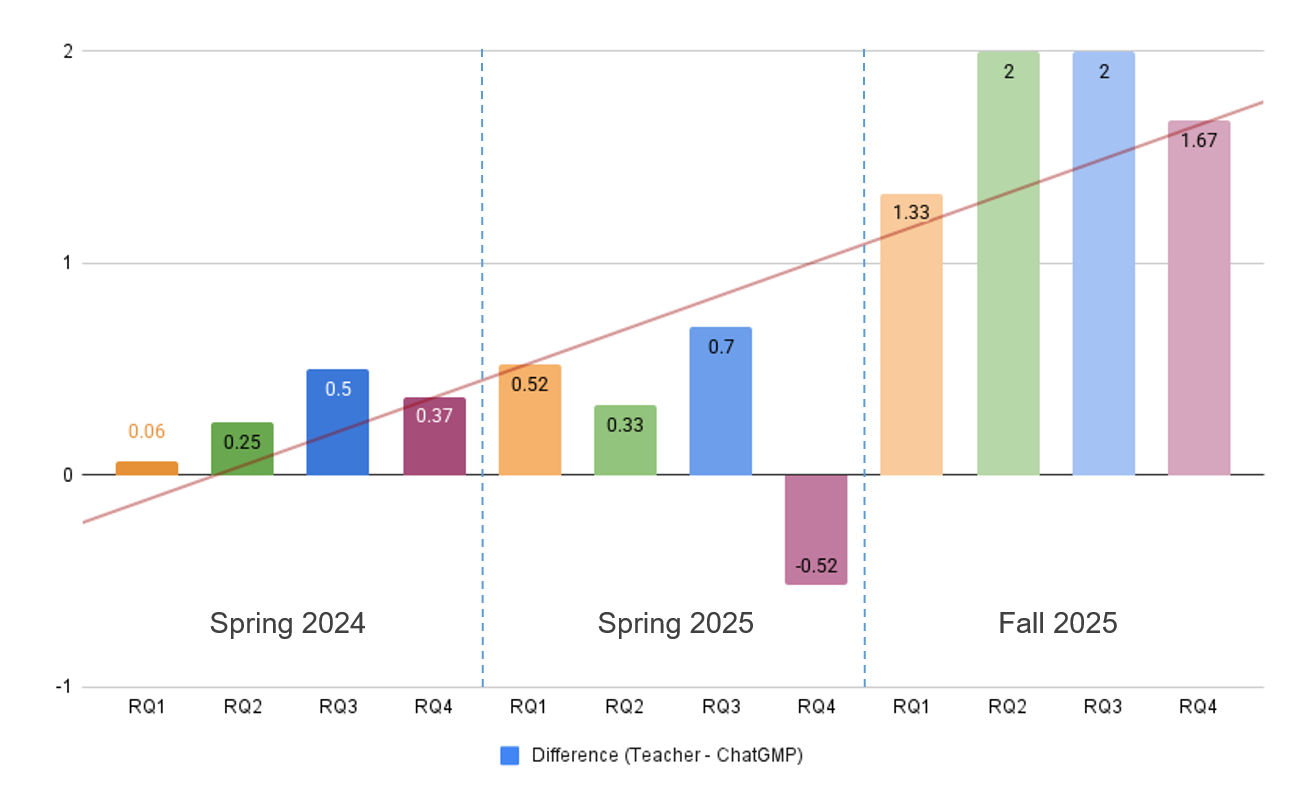}
\caption{\label{fig:difference}Difference between the average evaluation for the four research questions given to the teacher and AI assistant as auditee. The difference is positive if students evaluated higher the experience with the teacher, and negative if they preferred the AI assistant.}
\end{figure*}

Overall, students participating in the course in the spring semesters of 2024 and 2025 were happy with the tool and the quality of the answers, where an increase in satisfaction was observed in 2025. This is also evident in Figure \ref{fig:difference}, where more students think that performing the exercise with the AI assistant will be the future of the course.
On the other hand, the results presented in Table \ref{tab:performance} highlight an issue that deserves more attention and further evaluation. Currently, we do not have enough data to decouple performance from AI usage. However, if there is any correlation or whether the choice of using AI is in itself an indication of performance should be further investigated. In this study, we conclude that although there is a tendency for better grades from students who performed the audit exercise with a teacher, the AI assistant provided the necessary information for the groups to succeed in the learning objectives of the exercise and in writing a good report.

The difference between the perceptions of the students who audit a teacher and the AI assistant is shown in Figure \ref{fig:difference}. Here, a positive bar means that the students' ratings favored the teacher, and negative that they preferred the chatbot. The graph shows that, while in the normal course (2024 and 2025), the difference is relatively small, the crossover study leads to more polarized opinions.  
This shift could be due to different reasons. First, the cohort of students is different. The chatbot was developed for a M.Sc. course according to its specific learning objectives. In this course, students have no previous background in the subject, and therefore the audit is their first practical exercise in this area. This group of students usually likes the novelty introduced by the exercise, and they find it to be an innovative and effective way to test their understanding of the subject (since the teacher or chatbot would not provide them with the right documents if they are unable to ask for it correctly or be specific enough). Although they receive information during the audit, the deeper learning happens afterwards, when students have time to analyze and discuss the documents and detect the non-conformities and errors.
On the other hand, the students performing the audit in the fall semester of 2025 attend a more practical and professional-oriented M.Sc. course, where, as part of their studies, they are also employed half-time in a company. They have previous knowledge of the subject and they also have conducted audits in real-life situations before, therefore the expectations and prior knowledge of this cohort is much higher, compared to the students attending the course in the spring semester. Even the teacher performing the audit admitted to have been more challenged by this group. This is evident from some of their feedback, as presented below. 
\begin{itemize}
    \item "I did not particularly like the part that this was only 45 minutes. I think it would have been nice to have more time with the teacher to discuss more in depth the questions as we would think so much about the time that the real life experience of this (such as have a more elaborate discussion about a particular question)."
    \item "Usually the documentation part is done separately from an audit interview as the auditees will have time to go through the documents received and ask additional questions. Maybe in the future it would be nice that the students learn the meaning behind correct documentation and how to see for example missing data, mistakes in reporting data etc. that they could identify and ask additional questions."
    \item "Overall, it is a very impressive attempt to use AI for an audit but for real life comparison I do believe that 45 minutes is far from enough. Of course we also had a lot of questions we wanted answered and we got carried away so we would not go fully through the answers while on a real life interview you cannot avoid that. A nice test would be also for example have 10-15 main questions and start with the teacher an interview. Note down all additional questions asked during that interview and then follow exactly the same interview with ChatGMP. Time should not be taken into consideration for the cases and could be one of the things for comparison as well. Then the pressure of time is taken away leaving more focus to the answers receive and then getting a more precise evaluation of the whole exercise."
    \item "Provide more realistic documents, so students judge evidence, not just narratives."
\end{itemize} 

In summary, the crossover study also leads to interesting results: although the AI assistant was not developed for the cohort chosen for this investigation, and therefore we cannot assume that the results will be comparable to the other groups (of the students enrolling in the spring semesters of 2024 and 2025), the feedback is quite polarized. Thus, it provides many insights into the students' experience that will be incorporated to improve the model. For example, since the students performing the crossover study already had experience with the subject taught and auditing in general, they were more critical towards the limitations of the AI assistant, citing the fact that it was not able to handle complex questions and the experience lacked human interaction. This led us to re-evaluate the approach we are currently using; the model will be further improved before being used again in the spring semester of 2026. More details on the improvements we intend to make are discussed in Section \ref{future}. 


\subsection{Maximizing correctness vs. generating power trade-off}
A key learning from these three experiments is that choosing a smaller (and therefore less powerful) model to prioritize correctness and reduce hallucinations might not be the best strategy going forward, given the constant release of better models and the increase ease with which students interact with these AI-based tools. When we first used the model in 2024, students were not as used to interacting with commercially available chatbots such as ChatGPT. They tried for fun, but it was not until more recently that they started to use them more consistently for educational purposes. In a previous study published in 2024\footnote{The study will be referenced upon acceptance to avoid including information about the authors}, when interviewed about the possibility of having an AI assistant in their course, a student mentioned that the LLM "should definitely be better than what ChatGPT is now, if the level would be the same, the quality and performance would be low, in my opinion". This might suggest that the students did not consider it reliable in its earlier versions. However, reflecting on their experience after the audit with the teacher in 2025, students often mentioned that the AI assistant was not as good as ChatGPT, highlighting its recent improvements.
Therefore, students nowadays seem to have become accustomed to interacting with LLMs and therefore might have higher expectations regarding the quality of the answers, the speed, and the general language. 

However, this does not necessarily mean better quality of learning. More powerful LLMs might give the illusion of better knowledge provided, since the language might be more fluent than in previous models and the amount of information provided might be more extensive. However, hallucinations still pose a great risk in, among others, educational settings. 
Therefore, more research should be done on the safe and correct use of LLMs in education, discussing the learning outcomes of choosing to adopt smaller models that provide only correct information, even if not properly pertinent to the asked question (as done in this study) or more powerful models, with a higher risk of hallucinations.

\subsection{LLMs for teaching and learning: challenges, opportunities, and future directions}
There are many challenges hindering the widespread adoption of LLM-based educational technologies, both regarding teaching and learning. However, these challenges can lead to as many opportunities and future directions. We expand on these in teaching and learning by combining the key takeaways gained from this study with the background provided in Section \ref{background}. 
Table \ref{tab:challenges_opportunities_teaching} shows what we believe to be the main challenges of using LLMs in teaching, followed, respectively, by the opportunities they introduce and the potential future directions they open up. Table \ref{tab:challenges_opportunities_learning} uses the same approach, but focuses on the challenges, opportunities, and future directions of LLMs in learning. 

Among the challenges concerning teaching, \cite{yan2024practical} discuss the need for a human-centered approach in the design and implementation of these technologies. Furthermore, they stress that the majority of LLM-based innovations lack transparency beyond AI researchers, with minimal involvement of educational stakeholders in their development and evaluation. Additionally, they argue that LLMs generally demonstrate high performance in simpler classification tasks and promising results in prediction and generation, while their effectiveness in complex educational scenarios is still developing. 
There is a need to update the curricula to reflect new additions and include AI education, redesign assignments to promote higher-order thinking, and to train teachers to be able to face the new challenges introduced by AI \citep{Zhang2024Reflections, Idris2024Revolutionizing, caccavale2024llm}. Another concern for teaching includes the lack of reproducible studies, the fact that state-of-the-art (SOTA) research might not be applicable to the particular educational context and low technological readiness of universities.

A major challenge that should be further investigated is the effect that LLMs have on learning outcomes and the potential consequences of overdependence \citep{abd2023large}. In fact, as also highlighted by \cite{xu2024opportunities}, over-reliance on LLMs could harm learning by hindering critical thinking and reasoning skills. As discussed in Section \ref{background}, ethical concerns, transparency, privacy, and accountability are fundamental issues to be addressed and regulated before using LLMs in education, as they could have a negative impact on learning and potentially inhibit students' rights \citep{yan2024practical}. On the other hand, it is also important to educate students on the safe and ethical use of LLMs in order to prevent them from bending the rules of academic integrity, since these tools introduce the potential for plagiarism and cheating \citep{Perkins2023Academic, Kovari2025Ethical}. Other issues which needed to be addressed to safeguard learning concern equality and accessibility, given that the financial burden of LLMs and the limited access to infrastructures and technology, could aggravate inequity of learning and access for various student demographics \citep{Jin2023Better, Navigli2023BiasesIL}. Finally, the dominance of English-based models could exacerbate inequalities in language proficiency and accessibility \citep{Helm2024DiversityAL}.

\begin{table}[]
\centering
\caption{Challenges, opportunities, and future directions of LLMs in teaching.}
\label{tab:challenges_opportunities_teaching}
\begin{tabular}{|p{0.3\columnwidth}|p{0.3\columnwidth}|p{0.3\columnwidth}|} \hline
\multicolumn{3}{c}{\textbf{Teaching}} \\ \hline
\textbf{Challenges} & \textbf{Opportunities} & \textbf{Future directions}  \\ \hline
\item LLMs are continuously improving, so keeping up with AI developments can be hard for educational research & Constant improvement of LLMs means that increasingly more tasks (text-to-speech, video generation) could be automated with AI & Educators should recurrently set some time to investigate new AI developments and how they can be used to improve teaching \\ \hline
Ethical concerns, transparency, privacy and accountability are mandatory issues to be addressed before rolling out LLMs in educational contexts \citep{yan2024practical} & LLMs can provide help with a number of tasks, e.g.,  identify content gaps, suggest learning objectives, help with creating new content, provide help with assessment and evaluation & Establish clear guidelines for using LLMs in education, addressing ethical concerns, accountability, privacy and academic integrity, towards human-centric AI \citep{yan2024practical} \\ \hline 
Integrating AI  and other innovative technologies into education is a complex challenge learning objectives and teaching activities needs meticulous adjustments \citep{obidovna2024pedagogical, aggarwal2023integration} & Redesign the curriculum to reflect current advancements and leverage new technologies, automating repetitive tasks, increasing engagement and saving time & Update curricula to reflect new additions and include AI education, redesign assignments to promote higher-order thinking. Train teachers on ethics in AI \citep{Zhang2024Reflections, Idris2024Revolutionizing, caccavale2024llm}  \\ \hline 
Accuracy and reliability are major limitations of LLMs. Ensuring correctness of responses can be difficult due to the uncertainty of the information \citep{Augenstein2024Factuality} & Extract foundational knowledge of LLMs through prompt engineering. Investigate uncertainty in LLMs for deeper understanding & Use iterative approach to rigorously validate the models, extract uncertainty of generations and increase explainability and fairness \\ \hline 
Lack of replicability, as many studies fail to provide sufficient details and open-source code and data, making it difficult for educators to validate and adapt these innovations & Many libraries and tutorials on how to use LLMs in general scenarios can lead to the democratization of LLMs, making it accessible for non-AI experts to leverage these models in educational settings & Community should share code and data or, if not possible, provide detailed explanations on how to replicate the work and establish similar procedures in own applications \\ \hline
Published research and SOTA might not generalize to other pedagogical applications, due to cultural, institutional and technological differences of the specific university ecosystem & Collecting and analyzing a variety of opinions can lead to better understanding of social differences, if data is correctly analyzed & Perform cross-university/ country research to understand generalizable aspects and isolate culture-specific elements \\ \hline
Often analyses used to evaluate LLMs uses biased methods or subjective judgments, which can introduce variability in the results due to individual biases & Investigate how different factors (e.g., native language, culture) affect AI perception and proficiency. This can limit biases and tailor LLMs to specific needs and cultures & Focus on developing more robust and well-established metrics, how to benchmark LLMs and limit subjective biases \\ \hline
Most innovations remain in early development \citep{yan2024practical} & Perform new research that aims to contribute to the establishment of LLMs in education & Establish ways-of-working on how to embed and validate LLMs in authentic educational settings \\ \hline
Generational gap between educators and students might lead to suboptimal design decisions & Involving students as co-designers of the tools can lead to overall better experience and learning improvement
& Co-participatory design experiments to involve students in the implementation of LLMs
\\ \hline
\end{tabular}
\end{table}

\begin{table}[]
\centering
\caption{Challenges, opportunities, and future directions of LLMs in learning.}
\label{tab:challenges_opportunities_learning}
\begin{tabular}{|p{0.3\columnwidth}|p{0.3\columnwidth}|p{0.3\columnwidth}|} \hline
\multicolumn{3}{c}{\textbf{Learning}} \\ \hline
\textbf{Challenges} & \textbf{Opportunities} & \textbf{Future directions}  \\ \hline
Democratization and increasingly spread use of AI tools could lead to more individualized learning, since students might start to rely more on AI than on teachers and peers & Students can use LLMs to find more content related to their studies, find tutorials and additional material and in general find more accessible content & Provide assistance to students on how to safely use AI and embed it in group exercises to help students reflect on benefits and limitations to be addressed through discussions in the classroom \\ \hline
Over-reliance on LLMs could harm learning by hindering critical thinking and reasoning skills \citep{abd2023large, xu2024opportunities} & Ability to use LLMs in various learning settings, such as to quickly simplify complex information, provide prompt feedback & Teach students how to use LLMs correctly, emphasizing critical evaluation of AI-generated content and responsible usage \citep{nguyen2025use} \\ \hline
AI experience might feel dehumanizing for students, if they can only chat with a bot without personalization or human interaction & LLMs can be used to engage students through virtual simulations, interactive experience, and to enable personalized study plans and learning materials tailored to individual student needs & When using LLMs in classroom, plan more stimulant scenarios and simulations so that experience does not feel dehumanizing or impersonal \\ \hline 
LLMs might facilitate students to bend the rules of academic integrity, introducing potential for plagiarism and cheating & Access to information, summary and generation of ideas is much more available to students, as well writing assistance  & Provide students with clear guidelines on how to responsibly use LLMs in their coursework, addressing ethical and moral concerns, accountability and academic integrity \\ \hline 
Privacy issues are largely unaddressed, particularly concerning informed consent and data protection when fine-tuning LLMs with student data \citep{nguyen2023ethical} & Personalization of learning could lead to deeper knowledge gain and improved performance & When using LLMs with students, be transparent on what data is collected, who collects it and how it is used \\ \hline 
Financial burden of commercial LLMs, as well as limited access to infrastructures and technology, could aggravate inequity of learning and access for various student demographics & Development of distilled LLMs could facilitate and fasten learning in low income Countries & Distilled models to be deployed in lower income, less technology proficient Countries \\ \hline
The dominance of English-based models could exacerbate inequalities in language proficiency and accessibility & Understand how native speakers of different languages learn and if there are substantial differences & Increased focus on multilingual and monolingual languages models for non-English languages \\ \hline
\end{tabular}
\end{table}

\section{Future perspectives}
\label{future}
The underlying pre-trained LLM will be upgraded to one of the latest released models, to improve the quality of the generated responses and therefore the overall user experience.
More data will be curated and added to the dataset, improving the current background knowledge of the model.

One of the most distinctive caveats of this study is the lack of human interaction when using the AI assistant. To address this issue, next releases of the model will also include the development of a desktop and virtual reality experience where students will be able to interact with an avatar, giving the possibility to students to choose between the avatar and the chatbot medium.


\section{Limitations}
This study is subject to several potential limitations. A summary of identified limitations is as follows: (i) the analysis draws on a relatively small sample; (ii) the findings may have limited generalizability because they are tied to a specific teaching context; and, (iii) institutional structures and the particular technological ecosystem of the university where the study took place may constrain the extent to which these results can be transferred to other educational settings. 

Furthermore, noteworthy is that rapid advances in generative chatbots and evolving educational policies may influence the future relevance and validity of the results presented in this work. We recognize that the LLM used in this study is not the latest and most powerful and therefore has architectural constraints. Many models introduced in late 2025 can generate higher-quality outputs and are overall more capable. However, our choice of model was constrained by project resources, including both budget and GPU availability, and thus represented a reasonable compromise between fast inference and the provision of reliable and safe responses for students. 

Finally, the analysis relied on subjective judgments, which can introduce variability in the results due to personal opinions or preferences, as well as individual biases and levels of expertise. This inherent subjectivity underscores the breadth of possible perspectives, revealing diverse viewpoints and prompting researchers to consider their potential implications.

\section{Conclusions}
\label{conclusions}
This study examined a customized LLM-based assistant in an M.Sc. course. The model was validated across several experiments employing a mixed-methods iterative approach across three distinct experimental setups (spring 2024, spring 2025, and fall 2025 crossover study). We first compare the performance of the LLM between the spring semester of 2024, when it was initially introduced into coursework, and the spring semester of 2025. After integrating a portion of the feedback gathered during this period, the model was refined and subsequently assessed in greater depth through a crossover study conducted in the fall semester of 2025.

The results provide significant insights into the integration of LLMs in Higher Education, particularly in collaborative learning settings.
Our research highlights several critical trends regarding LLM use in education. Students were primarily motivated by curiosity and the desire to use an innovative tool. We found that they viewed the AI interaction as less stressful than interviewing a human teacher, which is a key psychological finding. Practical benefits also strongly drove adoption, including receiving a written conversation transcript and the AI's role in removing course enrollment limits. These findings suggest that addressing psychological barriers and offering concrete utility are powerful levers for LLM adoption in education.

Overall student satisfaction with the AI assistant and the quality of its answers increased from 2024 to 2025, a trend that aligns with the growing perception that the AI assistant represents the future of the course exercise.
A significant finding is the dramatic shift in student expectations regarding LLM performance. Students in later cohorts (2025) began to compare our custom AI assistant against high-performance commercial models (like ChatGPT), finding the custom model lacking in fluency and complexity. This indicates that the required quality threshold for educational LLM tools is constantly being elevated by external technological advancements.
The crossover study, involving more experienced students, yielded polarized and highly critical feedback. These students, possessing prior experience in the subject, exposed the limitations of the chatbot, citing its inability to handle complex questions and the overall lack of human interaction. This critical input was invaluable for identifying specific weaknesses and will guide our model refinement.

The study also identified a crucial pedagogical trade-off between prioritizing smaller, correctness-focused models (which may disappoint users due to lower fluency and limited scope, as we observed) and larger, more powerful models (which satisfy user expectations but carry a higher risk of hallucinations).

Future work will focus on model refinement and, crucially, investigating the learning outcomes associated with different LLM adoption strategies, specifically researching the safest and most effective use of LLMs in educational settings while minimizing the risk posed by hallucinations. We intend to further improve the model before using it again in the spring semester of 2026.

Finally, taking inspiration from our first-hand experience with implementing and testing LLMs in Higher Education, we discuss the challenges, opportunities, and future directions that we see in the foreseeable future of AI in education, both reflecting on its implications on teaching and on learning.

\section*{Statements on ethics}
Informed consent was obtained from all participants and their privacy rights were strictly observed. Before participating, all participants were informed that the collected data would be used for research purposes and would be handled anonymously. Detailed information on the study was provided to the participants before starting the survey, reassuring them of their right to voluntary participation and withdrawal. 

\section*{Availability of data and materials}
\label{code}
This study is committed to the open-source set of principles. The code is freely available on GitHub at \url{https://github.com/FiammettaC/ChatGMP}. Here, we also include a DEMO of the model that readers are free to test.
The code is published under the MIT license. Currently, it is compatible with Python 3.x, for Windows and Linux operating systems. It is also recommended to use a conda environment to run the code. Other data can be obtained by sending an email request to the corresponding author.

\section*{Funding}
Funding for this study was provided by (i) the European Union under Horizon Europe research and innovation program with the grant agreement 101159993, within the HORIZON-WIDERA-2023-ACCESS-02-0 call (Dig4Bio); (ii) Novo Nordisk Foundation, the grant number is: NNF22OC0080136 - Real-time sustainability analysis for Industry 4.0 (Sustain4.0); and (iii) Technical University of Denmark (DTU). 

\section*{Competing interests}
The authors have no competing interests to disclose.

\section*{Acknowledgments}
The authors thank the students who volunteered to test our AI assistant and helped us further develop the initiative.

\bigskip

\newpage

\bibliography{sn-bibliography}

@article{lewis2020retrieval,
  title={Retrieval-augmented generation for knowledge-intensive nlp tasks},
  author={Lewis, Patrick and Perez, Ethan and Piktus, Aleksandra and Petroni, Fabio and Karpukhin, Vladimir and Goyal, Naman and K{\"u}ttler, Heinrich and Lewis, Mike and Yih, Wen-tau and Rockt{\"a}schel, Tim and others},
  journal={Advances in neural information processing systems},
  volume={33},
  pages={9459--9474},
  year={2020}
}

@article{caccavale2024llm,
  title={Towards Education 4.0: The Role of Large Language Models as Virtual Tutors in Chemical Engineering},
  author={Caccavale, Fiammetta and Gargalo, Carina L and Gernaey, Krist V and Kr{\"u}hne, Ulrich},
  journal={Education for Chemical Engineers},
  volume={49},
  pages={1--11},
  year={2024},
  publisher={Elsevier}
}

@article{chung2024scaling,
  title={Scaling instruction-finetuned language models},
  author={Chung, Hyung Won and Hou, Le and Longpre, Shayne and Zoph, Barret and Tay, Yi and Fedus, William and Li, Yunxuan and Wang, Xuezhi and Dehghani, Mostafa and Brahma, Siddhartha and others},
  journal={Journal of Machine Learning Research},
  volume={25},
  number={70},
  pages={1--53},
  year={2024}
}

@article{yan2024practical,
  title={Practical and ethical challenges of large language models in education: A systematic scoping review},
  author={Yan, Lixiang and Sha, Lele and Zhao, Linxuan and Li, Yuheng and Martinez-Maldonado, Roberto and Chen, Guanliang and Li, Xinyu and Jin, Yueqiao and Ga{\v{s}}evi{\'c}, Dragan},
  journal={British Journal of Educational Technology},
  volume={55},
  number={1},
  pages={90--112},
  year={2024},
  publisher={Wiley Online Library}
}

@inproceedings{papineni2002bleu,
  title={Bleu: a method for automatic evaluation of machine translation},
  author={Papineni, Kishore and Roukos, Salim and Ward, Todd and Zhu, Wei-Jing},
  booktitle={Proceedings of the 40th annual meeting of the Association for Computational Linguistics},
  pages={311--318},
  year={2002}
}

@article{kasneci2023chatgpt,
  title={ChatGPT for good? On opportunities and challenges of large language models for education},
  author={Kasneci, Enkelejda and Se{\ss}ler, Kathrin and K{\"u}chemann, Stefan and Bannert, Maria and Dementieva, Daryna and Fischer, Frank and Gasser, Urs and Groh, Georg and G{\"u}nnemann, Stephan and H{\"u}llermeier, Eyke and others},
  journal={Learning and individual differences},
  volume={103},
  pages={102274},
  year={2023},
  publisher={Elsevier}
}

@article{kamalov2023new,
  title={New era of artificial intelligence in education: Towards a sustainable multifaceted revolution},
  author={Kamalov, Firuz and Santandreu Calonge, David and Gurrib, Ikhlaas},
  journal={Sustainability},
  volume={15},
  number={16},
  pages={12451},
  year={2023},
  publisher={MDPI}
}

@article{rodrigues2024assessing,
  title={Assessing the quality of automatic-generated short answers using GPT-4},
  author={Rodrigues, Luiz and Pereira, Filipe Dwan and Cabral, Luciano and Ga{\v{s}}evi{\'c}, Dragan and Ramalho, Geber and Mello, Rafael Ferreira},
  journal={Computers and Education: Artificial Intelligence},
  volume={7},
  pages={100248},
  year={2024},
  publisher={Elsevier}
}

@article{gkintoni2025challenging,
  title={Challenging cognitive load theory: The role of educational neuroscience and artificial intelligence in redefining learning efficacy},
  author={Gkintoni, Evgenia and Antonopoulou, Hera and Sortwell, Andrew and Halkiopoulos, Constantinos},
  journal={Brain Sciences},
  volume={15},
  number={2},
  pages={203},
  year={2025},
  publisher={MDPI}
}

@article{caccavale2025chatgmp,
  title={ChatGMP: A case of AI chatbots in chemical engineering education towards the automation of repetitive tasks},
  author={Caccavale, Fiammetta and Gargalo, Carina L and Kager, Julian and Larsen, Steen and Gernaey, Krist V and Kr{\"u}hne, Ulrich},
  journal={Computers and Education: Artificial Intelligence},
  volume={8},
  pages={100354},
  year={2025},
  publisher={Elsevier}
}

@article{lavicza2022developing,
  title={Developing and evaluating educational innovations for STEAM education in rapidly changing digital technology environments},
  author={Lavicza, Zsolt and Weinhandl, Robert and Prodromou, Theodosia and An{\dj}i{\'c}, Branko and Lieban, Diego and Hohenwarter, Markus and Fenyvesi, Kristof and Brownell, Christopher and Diego-Mantec{\'o}n, Jose Manuel},
  journal={Sustainability},
  volume={14},
  number={12},
  pages={7237},
  year={2022},
  publisher={MDPI}
}

@article{obidovna2024pedagogical,
  title={The pedagogical-psychological aspects of artificial intelligence technologies in integrative education},
  author={Obidovna, Djalilova Zarnigor},
  journal={International Journal Of Literature And Languages},
  volume={4},
  number={03},
  pages={13--19},
  year={2024}
}

@article{aggarwal2023integration,
  title={Integration of innovative technological developments and AI with education for an adaptive learning pedagogy},
  author={Aggarwal, Deepshikha},
  journal={China Petroleum Processing and Petrochemical Technology},
  volume={23},
  number={2},
  pages={709--714},
  year={2023}
}

@article{chiu2024future,
  title={Future research recommendations for transforming higher education with generative AI},
  author={Chiu, Thomas KF},
  journal={Computers and education: Artificial intelligence},
  volume={6},
  pages={100197},
  year={2024},
  publisher={Elsevier}
}

@article{ng2023teachers,
  title={Teachers’ AI digital competencies and twenty-first century skills in the post-pandemic world},
  author={Ng, Davy Tsz Kit and Leung, Jac Ka Lok and Su, Jiahong and Ng, Ross Chi Wui and Chu, Samuel Kai Wah},
  journal={Educational technology research and development},
  volume={71},
  number={1},
  pages={137--161},
  year={2023},
  publisher={Springer}
}

@article{nguyen2023ethical,
  title={Ethical principles for artificial intelligence in education},
  author={Nguyen, Andy and Ngo, Ha Ngan and Hong, Yvonne and Dang, Belle and Nguyen, Bich-Phuong Thi},
  journal={Education and information technologies},
  volume={28},
  number={4},
  pages={4221--4241},
  year={2023},
  publisher={Springer}
}

@article{kelly2023chatgpt,
  title={ChatGPT in higher education: Considerations for academic integrity and student learning},
  author={Kelly, Andrew and Sullivan, Miriam},
  journal={Journal of applied learning and teaching},
  year={2023}
}

@article{nguyen2025use,
  title={The use of generative AI tools in higher education: Ethical and pedagogical principles},
  author={Nguyen, Khoa Viet},
  journal={Journal of Academic Ethics},
  pages={1--21},
  year={2025},
  publisher={Springer}
}

@misc{huggingface,
  author = {HuggingFace},
  title = {{Hugging Face}},
  howpublished = "\url{https://huggingface.co/}",
  year = {2016}, 
  note = "[Online; accessed 09-October-2025]"
}

@article{mikolov2013distributed,
  title={Distributed representations of words and phrases and their compositionality},
  author={Mikolov, Tomas and Sutskever, Ilya and Chen, Kai and Corrado, Greg S and Dean, Jeff},
  journal={Advances in neural information processing systems},
  volume={26},
  year={2013}
}

@article{mann1947test,
  title={On a test of whether one of two random variables is stochastically larger than the other},
  author={Mann, Henry B and Whitney, Donald R},
  journal={The annals of mathematical statistics},
  pages={50--60},
  year={1947},
  publisher={JSTOR}
}

@article{mchugh2013chi,
  title={The chi-square test of independence},
  author={McHugh, Mary L},
  journal={Biochemia medica},
  volume={23},
  number={2},
  pages={143--149},
  year={2013},
  publisher={Medicinska naklada}
}

@misc{gmp_grades,
  author = {DTU.dk},
  title = {{28855 GMP og kvalitet i farmaceutisk, biotek og fødevareindustri - teoretisk version, Sommer 2025}},
  howpublished = "\url{https://karakterer.dtu.dk/Histogram/1/28855/Summer-2025}",
  year = {2025}, 
  note = "[Online; accessed 29-October-2025]"
}

@misc{danish_grades,
  author = {DTU.dk},
  title = {{Grading system of Danish higher
education}},
  howpublished = "\url{https://studyabroad.dtu.dk/english/-/media/subsites/study_abroad/dokumenter/karakterer_videregaaende_en.pdf}",
  year = {2025}, 
  note = "[Online; accessed 29-October-2025]"
}

@article{Gan2023Large,
  title={Large Language Models in Education: Vision and Opportunities},
  author={Wensheng Gan and Zhenlian Qi and Jiayang Wu and Chun-Wei Lin},
  journal={2023 IEEE International Conference on Big Data (BigData)},
  year={2023},
  pages={4776-4785},
  doi={10.1109/bigdata59044.2023.10386291}
}

@article{Xu2024Large,
  title={Large Language Models for Education: A Survey},
  author={Hanyi Xu and Wensheng Gan and Zhenlian Qi and Jiayang Wu and Philip S. Yu},
  journal={ArXiv},
  year={2024},
  volume={abs/2405.13001},
  doi={10.48550/arxiv.2405.13001}
}

@article{Pelaez-Sanchez2024The,
  title={The impact of large language models on higher education: exploring the connection between AI and Education 4.0},
  author={Iris Cristina Pelaez-Sanchez and Davis Velarde-Camaqui and Leonardo-David Glasserman-Morales},
  journal={Frontiers in Education},
  year={2024},
  doi={10.3389/feduc.2024.1392091}
}

@article{Zhang2024Reflections,
  title={Reflections on Enhancing Higher Education Classroom Effectiveness Through the Introduction of Large Language Models},
  author={Xiaoming Zhang and Xiaoli Zhang and He Liu},
  journal={Journal of Modern Educational Research},
  year={2024},
  doi={10.53964/jmer.2024019}
}

@article{Idris2024Revolutionizing,
  title={Revolutionizing Higher Education: Unleashing the Potential of Large Language Models for Strategic Transformation},
  author={Mohamed Diab Idris and Xiaohua Feng and Vladimir Dyo},
  journal={IEEE Access},
  year={2024},
  volume={12},
  pages={67738-67757},
  doi={10.1109/access.2024.3400164}
}

@article{naik2025providing,
  title={Providing tailored reflection instructions in collaborative learning using large language models},
  author={Naik, Atharva and Yin, Jessica Ruhan and Kamath, Anusha and Ma, Qianou and Wu, Sherry Tongshuang and Murray, R Charles and Bogart, Christopher and Sakr, Majd and Rose, Carolyn P},
  journal={British Journal of Educational Technology},
  volume={56},
  number={2},
  pages={531--550},
  year={2025},
  publisher={Wiley Online Library}
}

@inproceedings{naik2024generating,
  title={Generating situated reflection triggers about alternative solution paths: A case study of generative ai for computer-supported collaborative learning},
  author={Naik, Atharva and Yin, Jessica Ruhan and Kamath, Anusha and Ma, Qianou and Wu, Sherry Tongshuang and Murray, Charles and Bogart, Christopher and Sakr, Majd and Rose, Carolyn P},
  booktitle={International Conference on Artificial Intelligence in Education},
  pages={46--59},
  year={2024},
  organization={Springer}
}

@inproceedings{cai2024advancing,
  title={Advancing knowledge together: integrating large language model-based conversational AI in small group collaborative learning},
  author={Cai, Zhenyao and Park, Seehee and Nixon, Nia and Doroudi, Shayan},
  booktitle={Extended Abstracts of the CHI Conference on Human Factors in Computing Systems},
  pages={1--9},
  year={2024}
}

@inproceedings{lewis2022multimodal,
  title={Multimodal large language models for inclusive collaboration learning tasks},
  author={Lewis, Armanda},
  booktitle={Proceedings of the 2022 Conference of the North American Chapter of the Association for Computational Linguistics: Human Language Technologies: Student Research Workshop},
  pages={202--210},
  year={2022}
}

@article{xu2024opportunities,
  title={Opportunities, challenges, and future directions of large language models, including ChatGPT in medical education: a systematic scoping review},
  author={Xu, Xiaojun and Chen, Yixiao and Miao, Jing},
  journal={Journal of educational evaluation for health professions},
  volume={21},
  year={2024},
  publisher={Korea Health Personnel Licensing Examination Institute}
}

@article{abd2023large,
  title={Large language models in medical education: opportunities, challenges, and future directions},
  author={Abd-Alrazaq, Alaa and AlSaad, Rawan and Alhuwail, Dari and Ahmed, Arfan and Healy, Padraig Mark and Latifi, Syed and Aziz, Sarah and Damseh, Rafat and Alrazak, Sadam Alabed and Sheikh, Javaid},
  journal={JMIR medical education},
  volume={9},
  number={1},
  pages={e48291},
  year={2023},
  publisher={JMIR Publications Inc., Toronto, Canada}
}

@article{Augenstein2024Factuality,
  title={Factuality challenges in the era of large language models and opportunities for fact-checking},
  author={Isabelle Augenstein and Timothy Baldwin and Meeyoung Cha and Tanmoy Chakraborty and Giovanni Luca Ciampaglia and David Corney and Renee DiResta and Emilio Ferrara and Scott Hale and A. Halevy and Eduard H. Hovy and Heng Ji and Filippo Menczer and Rubén Míguez and Preslav Nakov and Dietram A. Scheufele and Shivam Sharma and Giovanni Zagni},
  journal={Nat. Mac. Intell.},
  year={2024},
  volume={6},
  pages={852-863},
  doi={10.1038/s42256-024-00881-z}}

@article{Jin2023Better,title={Better to Ask in English: Cross-Lingual Evaluation of Large Language Models for Healthcare Queries},author={Yiqiao Jin and Mohit Chandra and Gaurav Verma and Yibo Hu and Munmun De Choudhury and Srijan Kumar},journal={Proceedings of the ACM Web Conference 2024},year={2023},doi={10.1145/3589334.3645643}}

@article{Navigli2023BiasesIL, author = {Roberto Navigli and Simone Conia and Björn Ross}, journal = {ACM Journal of Data and Information Quality}, title = {Biases in large language models: origins, inventory, and discussion}, year = {2023}, volume = {15}, number = {2}, pages = {1-21} }

@article{Helm2024DiversityAL, title={Diversity and language technology: how language modeling bias causes epistemic injustice}, author={Paula Helm and Gábor Bella and Gertraud Koch and Fausto Giunchiglia}, journal={Ethics and Information Technology}, volume={26}, number={1}, pages={8}, year={2024} }

@article{Perkins2023Academic,title={Academic integrity considerations of AI Large Language Models in the post-pandemic era: ChatGPT and beyond},author={Mike Perkins},journal={Journal of University Teaching and Learning Practice},year={2023},doi={10.53761/1.20.02.07}}

@article{Kovari2025Ethical,title={Ethical use of ChatGPT in education—Best practices to combat AI-induced plagiarism},author={Attila Kovari},journal={Frontiers in Education},year={2025},doi={10.3389/feduc.2024.1465703}}

\newpage
\appendix

\begin{table}[]
\centering
\caption{Average results of FLAN-T5 and LLaMA 2 models given the prompt (student's question) and context (teacher's answer). The results are calculated over a subset of questions (N=10) and A100 GPU was used. The metrics reported are the cosine similarity and BLEU score. Cosine measures, such as similarity and distance, are commonly used in NLP to calculate the relationship between words, sentences or documents, as in \cite{mikolov2013distributed}. BLEU is a precision-based metric originally developed for evaluating machine translation and it quantifies alignment between the generated and the reference text \citep{papineni2002bleu}. Adapted from **Anonymized** (reference will be added upon acceptance).}
\label{tab:results_appendix}
\begin{tabular}{lccc}
                  & \textbf{Cosine similarity} & \textbf{BLEU}   & \textbf{Runtime} \\ \hline

\textbf{FLAN-T5 base} & 0.68 & 0.39 & 19.0 s                      \\ \hline
\textbf{LLaMA 2} & 0.72 & 0.33 & 38.9 s               \\ \hline
\end{tabular}
\end{table}

\begin{table}[]
\centering
\caption{Students' perspectives regarding the quality of the answers provided by the AI assistant.}
\label{tab:feedback_RQ2}
\begin{tabular}
{p{0.95\columnwidth}}
\textbf{RQ2: What do you think of the quality of the answers?} \\ \hline
\begin{itemize}
    \item Some of the answers were the same, though i think this was our fault as we found prompting hard.
    \item It felt like it had a set of pre-made answers and couldn't really elaborate as if it was a conversation. Many times the answers were unrelated to the question
    \item You may structure it in such a way the ChatGMP asks the auditor to be explicit with the question when it doesn't understand the question rather than giving "Yes, of course"
    \item The questions need to be very specific, and sometimes it would take time to reformulate the question so we could get the expected answer. In many cases we did not get what we wanted.
    \item For most of the answers yes it was really nice, but for few ones it didn't make sense to us. If it was a real audit in person then it wouldn't be acceptable
    \item When you actually get what you are asking for, then it is good, but because it was so difficult for the chat to understand the questions, the answers was bad
    \item Live person elaborate more than bot
    \item It felt like nuance is lost, slight changes to a question to hone in on specific answers would yield the same response. Other than that it seemed like the responses to prompts were given character to simulate a human - which made it more personal.
    \item The transcript is very convenient, making it easy to scan for information from the answers.
    \item While it was really cool that the answers were good and were more similar to a real person talking rather than ChatGPT, it was really difficult to read the answers during the audit as the text was formulated like speech.
    \item If our questions were too specific, the bot got very confused.
    \item In general, the answers were very good. The LLM has been trained well on the topic, and provides very coherent and professional answers. Sometimes we wondered if the chatbot was even more honest than a real person would have been about the issues in Pharma A/S. In a few cases, I think 2 questions, the chatbot had issues understanding what we were asking for. It seems that it might have been trained a bit rigidly, considering that there is a difference between asking for X program or X procedure.
    \item We had to be really specific on the questions and be precise with the words we were using.
    \item Sometimes the answers were completely unrelated but that is to be expected as the virtual tutor is still learning.
    \item The quality of the answers were ok, but it was quite hard to ask the questions in a way so that the bot would understand what we were looking for.
    \item The answers seem really real, feel like doing audit with real companies. Helped us be more familiar with audit processes and also improve our questions list.
    \item not always clear
    \item Overall it was a very good experience. it was very honest and the answers were of appropriate length. It didn't understand some of our questions, when we didn't use the right words, such as using program instead of protocol. but just asking it again, fixed the issues. Sometimes we asked questions, which it didn't have an answer to or understand, but it most likely would have been the same with an audit person.
\end{itemize} \\ \hline
\end{tabular}
\end{table}

\begin{table}[]
\centering
\caption{Students' perspectives regarding the quality of the answers provided by the teacher in the spring semester of 2025.}
\label{tab:feedback_RQ2_teachers}
\begin{tabular}
{p{0.95\columnwidth}}
\textbf{RQ2: What do you think of the quality of the answers?} \\ \hline
\begin{itemize}
    \item Very close to reality i Think 
    \item Answers were designed to highlight mistakes clearly
    \item Most of them were answered and documentation was supplied, although it was available a bit later than we expected. 
    \item Teacher had answers for most of our questions and could refer to the documents
    \item I was a bit confused about how to handle answers that talked about a document that was not provided. I understand that not all QMS documents for NovoPharma can be created, but it would have been nice to have been explained how to handle documents, that the auditee said "existed" but could not provide. 
    \item The teacher did try to talk about some random things, but that was expected so we focused on the relevant answers. The documents that we recieved after were very good. 
    \item Also mock instalations could be added as it makes it more visual and helps the students to have a notion of visual inspections and keeping an eye for details. Maybe high res pictures that you can zoom in to see if you find any deviation, or videos with normal/abnormal procedures. It is also more engaging rather than just doing a documentative review/control.
    \item I understand that the teacher needs to speak for "gain time" for their company, but I felt bad/rude when I cut his conversation. But I know that he is simulating a real life. 
    \item The quality of the answers were good. They didnt have all the documents that we asked but its not a real world audit so its fine.
    \item It was okay
    \item Sometimes missing info but very reasonable
    \item I actually also learned something about working with GMP and about procedures in pharmaceutical production during the actual audit. 
    \item I think we got accurate answers to most of our questions. There were some momwnts that we were asking for some evidence and non was provided, and I wasn's sure if this was just because it was a fake audit, or I should have registered it as a finding. 
\end{itemize} \\ \hline
\end{tabular}
\end{table}

\begin{table}[]
\centering
\caption{Students’ perspectives regarding the future of the exercise. Responses provided by students that audited a teacher in the spring semester of 2025.}
\label{tab:future_teacher}
\begin{tabular}
{p{0.95\columnwidth}}
\textbf{RQ4: Do you think this is the future of the audit exercise in this course?} \\ \hline
\begin{itemize}
    \item The closest experience you can have to the actual job, offering a valuable complement to traditional lectures
    \item It was an overall good experience with the teacher, don't really have a comparison. For me it came to a certain extent close to a real audit and I would recommend others to do it with a teacher as well. 
    \item It forces you to create a list of questions and have a critical view of the answers/documents you receive. 
    \item I think the exercise worked very well
    \item The AI assistant has a lot of potential the more refined it gets. But It will never beat a good acted mock audit with a human auditee
    \item I think is a real and good exercise to prepare us even if we want or not to work in QA or audit department. you can learn a lot and is useful for your future profesional life
    \item Yes, it was nice to create a real world audit , so we can learn how the audit process goes. There was sufficient time for each team work. If the guest lectures from food industry would be nice too
    \item Not sure
    \item I didnt do it with chat gmp 
    \item I havn't tried the chat version
    \item Personally, I have never been part of an audit before, and I think that this exercise gave me a good representation of what audits are about. the departments that are involved and a general image of the procedure. 
\end{itemize} \\ \hline
\end{tabular}
\end{table}

\begin{table}[]
\centering
\caption{Additional reflections of the students regarding the interaction with the AI assistant.}
\label{tab:feedback_additional_info}
\begin{tabular}
{p{0.95\columnwidth}}
\textbf{Additional feedback} \\ \hline
\begin{itemize}
    \item It was harder to prompt than what i am used to, as it did not have memory of the last question, so follow up questions was seemless.
    \item It felt very planned and it looks like the model was just trained with a specific sets of answers and was just trying to find a best fit for the questions
    \item It was very nice to try this out, it felt more relaxed and in control of the next steps. Even though, in some cases it was difficult to receive the answers we wanted, or the documents, it was still nice to work with it.
    \item It was fun to try
    \item Clarifying questions were not the easiest, and with the machine sometimes producing identical answers it did sound a little robotic, for lack of a better word.
    \item Text was very uninviting to read in general. The slow responses made it not feel more like a computer than a "real company"
    \item It was very communicating in a very professional manner, so except for 2 times where it misunderstood our questions, I had the impression that we could be talking to a real person.
    \item We spent some time (too much time in my opinion) rephrasing our questions to the bot several times because it didn't understand what we meant, this made me quite stressed. And I feel like sometimes our question was too narrow and detailed, but sometimes too broad and so I had a hard time getting the hang of how to phrase the questions.
    \item It’s really nice but maybe would be better if we could copy and paste our questions instead of typing.
    \item still early stage development AI
    \item It was a more relaxed and postive experience interacting with the chat, which was nice to take the pressure off.
    \item It was particularly good at providing documents.
\end{itemize}
\\ \hline
\end{tabular}
\end{table}

\end{document}